\documentclass{ifacconf}

\usepackage{algorithm}
\usepackage{amsmath}
\usepackage{amssymb}
\usepackage{algpseudocode}

\usepackage{array}
\newcolumntype{L}[1]{>{\raggedright\arraybackslash}p{#1}}   
\newcolumntype{C}[1]{>{\centering\arraybackslash}p{#1}}     
\newcolumntype{R}[1]{>{\raggedleft\arraybackslash}p{#1}}    
\usepackage{bm}

\usepackage{graphicx}      

\usepackage{mathrsfs}
\usepackage{mathtools}

\usepackage[square,authoryear]{natbib}

\usepackage{pgfplots}
\pgfplotsset{compat=1.18}
\usepackage{subcaption}
\usetikzlibrary{external}

\usepackage{multirow} 

\usepackage[bookmarks=true,hidelinks]{hyperref}
\usepackage[acronym]{glossaries}

\usepackage{tikz}
\usepackage{titlesec}
\AtBeginDocument{\setcounter{secnumdepth}{3}}

\titleformat{\subsubsection}[block]{\normalfont\itshape}{\thesubsubsection.}{0.5em}{}
\titlespacing*{\subsubsection}{0pt}{1ex plus .2ex}{0.8ex} 

\AtBeginDocument{}
\AtBeginDocument{}
\AtBeginDocument{}

\newtheorem{theorem}{Theorem}

\newtheorem{definition}{Definition}

\newtheorem{lemma}{Lemma}

\newtheorem{corollary}{Corollary}

\DeclarePairedDelimiter\set{\{}{\}}

\newcommand{\col}{\text{col}}

\newcommand{\diag}{\text{diag}}
\newcommand{\tr}{\text{Trace}}
\newcommand{\st}{\text{s.t.}}

\newcommand{\he}{\text{He}}

\newcommand{\norm}[1]{\left\| #1 \right\|}

\newcommand{\brackets}[1]{\left( #1 \right)}

\newcommand{\quadsupplyinput}[5]{\begin{bmatrix}
    #4 \\ #5
\end{bmatrix}^{\hspace{-1pt}T}\hspace{-4pt}\begin{bmatrix}
    #1 & #2 \\ * & #3
\end{bmatrix}\hspace{-3pt}\begin{bmatrix}
    #4 \\ #5
\end{bmatrix}}

\newcommand{\dbx}{\dot{\textbf{x}}}

\newcommand{\cC}{\mathcal{C}}

\newcommand{\cG}{\mathcal{G}}
\newcommand{\cH}{\mathcal{H}}

\newcommand{\cK}{\mathcal{K}}
\newcommand{\cL}{\mathcal{L}}

\newcommand{\cX}{\mathcal{X}}

\newcommand{\bbN}{\mathbb{N}}

\newcommand{\bbR}{\mathbb{R}}
\newcommand{\bbS}{\mathbb{S}}

\newcommand{\bzero}{\mathbf{0}}

\newcommand{\bA}{\mathbf{A}}
\newcommand{\bB}{\mathbf{B}}
\newcommand{\bC}{\mathbf{C}}
\newcommand{\bD}{\mathbf{D}}

\newcommand{\bF}{\mathbf{F}}
\newcommand{\bG}{\mathbf{G}}
\newcommand{\bH}{\mathbf{H}}
\newcommand{\bI}{\mathbf{I}}

\newcommand{\bK}{\mathbf{K}}
\newcommand{\bL}{\mathbf{L}}
\newcommand{\bM}{\mathbf{M}}
\newcommand{\bN}{\mathbf{N}}

\newcommand{\bP}{\mathbf{P}}
\newcommand{\bQ}{\mathbf{Q}}
\newcommand{\bR}{\mathbf{R}}
\newcommand{\bS}{\mathbf{S}}

\newcommand{\bV}{\mathbf{V}}

\newcommand{\bX}{\mathbf{X}}
\newcommand{\bY}{\mathbf{Y}}
\newcommand{\bZ}{\mathbf{Z}}

\newcommand{\be}{\mathbf{e}}

\newcommand{\br}{\mathbf{r}}

\newcommand{\bu}{\mathbf{u}}

\newcommand{\bx}{\mathbf{x}}
\newcommand{\by}{\mathbf{y}}

\newcommand{\scrC}{\mathscr{C}}

\newcommand{\scrG}{\mathscr{G}}
\newcommand{\scrH}{\mathscr{H}}

\newcommand{\hatbe}{\widehat{\be}}

\newcommand{\hatbr}{\widehat{\br}}

\newcommand{\hatbu}{\widehat{\bu}}

\newcommand{\hatby}{\widehat{\by}}

\newcommand{\hatbA}{\widehat{\bA}}

\newcommand{\hatbQ}{\widehat{\bQ}}
\newcommand{\hatbR}{\widehat{\bR}}
\newcommand{\hatbS}{\widehat{\bS}}

\newcommand{\tilbH}{\widetilde{\bH}}

\newcommand{\thh}{\mathrm{th}}

\newacronym{vndt}{NDT}{Network Dissipativity Theorem}
\newacronym{io}{IO}{input-output}

\newacronym{lti}{LTI}{linear time-invariant}
\newacronym[plural=LMIs,longplural=linear matrix inequalities]{lmi}{LMI}{linear matrix inequality}
\newacronym{lqr}{LQR}{linear quadratic regulator}
\newacronym{lq}{LQ}{linear quadratic}

\newacronym{admm}{ADMM}{alternating direction methods of multipliers}
\newacronym{sdp}{SDP}{semidefinite programming}
\newacronym[plural=BMIs,longplural=bilinear matrix inequalities]{bmi}{BMI}{bilinear matrix inequality}
\newacronym{ico}{ICO}{iterative convex overbounding}

\newacronym{uav}{UAV}{unmanned aerial vehicles}
\newacronym{cps}{CPS}{cyber-physical system}
\def\equationautorefname#1#2\null{(#2\null)}

\newcommand{\dbG}{\delta{\bG}}
\newcommand{\dbX}{\delta{\bX}}
\newcommand{\dbY}{\delta{\bY}}
\newcommand{\dbK}{\delta{\bK}}
\newcommand{\dbH}{\delta{\bH}}
\newcommand{\dbQ}{\delta{\bQ}}
\newcommand{\dbS}{\delta{\bS}}
\newcommand{\dbR}{\delta{\bR}}
\newcommand{\dbP}{\delta{\bP}}
\newcommand{\hatdbQ}{\delta{\hatbQ}}

\newcommand{\card}{\text{card}}
\begin{document}
\sloppy
\begin{frontmatter}

\title{Communication-Aware Dissipative Control for Networks of Heterogeneous Nonlinear Agents} 

\thanks[footnoteinfo]{This work is supported by ONR Grant No. N00014-23-1-2043.}

\author[First]{Ingyu Jang} 
\author[First]{Leila J. Bridgeman} 

\address[First]{Department of Mechanical Engineering and Material Science, Duke University, Durham, NC 27708 USA (e-mail: ij40@duke.edu; ljb48@duke.edu).}

\begin{abstract}                
Communication-aware control is essential to reduce costs and complexity in large-scale networks.  
However, it is challenging to simultaneously determine a sparse communication topology and achieve high performance and robustness.
This work achieves all three objectives through dissipativity-based, sparsity-promoting controller synthesis. 
The approach identifies an optimal sparse structure using either weighted $\ell_1$ penalties or \gls{admm} with a cardinality term, and iteratively solves a convexified version of the NP hard structured optimal control problem. 
The proposed methods are demonstrated on heterogeneous networks with uncertain and unstable agents.
\end{abstract}


\end{frontmatter}

\section{Introduction}
Networked control systems arise in many modern applications, such as smart grids, power plants networks, or swarm robotics. 
Even though classical, fully-connected controllers can stabilize such networks with reasonable robustness, they impose substantial communication demands and often lead to impractical communication infrastructures. 
On the other hand, fully decentralized controllers, which require no communication between agents, eliminate communication costs entirely but typically yield significantly degraded performance.
Consequently, there is a strong need for controller architectures that balance performance and communication efficiency by reducing interconnections among agents while maintaining desirable closed-loop behavior \citep{jovanovic2016controller}. 
Various controller synthesis methods have been proposed to reduce the communication costs in networked control systems, but designing sparse controllers which are robust and applicable to general heterogeneous nonlinear agents remains a challenging problem.
This paper addresses the synthesis of robust, communication-aware, controllers for networks of heterogeneous nonlinear agents by promoting controller sparsity and assuring network-wide stability through the \gls{vndt}.

Energy-related analysis is widely accepted in the study of stability for physical and engineering systems.
Dissipativity \citep{willems1972dissipative,hill2003stability} provides an abstract, energy-based characterization of general nonlinear systems by treating them as input-output mappings, ignoring internal states. 
A key advantage of this abstraction is its design compositionality \citep{zakeri2022passivity}, meaning that the interconnections of dissipative systems are themselves dissipative and stable with minimum conditions.
The \gls{vndt} in \citep{moylan2003stability,vidyasagar1981input} exploits this to assess network-level stability using only the open-loop properties of individual agents.
While the theorem has been successfully applied to the synthesis of decentralized controllers for large-scale networks \citep{arcak2016networks,locicero2025dissipativity}, additional mechanisms are required when communication among controllers is necessary to improve the performance of the controller.

A straightforward approach to reducing controller interconnections is to impose an $\ell_0$ norm (cardinality) constraint on the optimal control framework.
However, optimization problems involving cardinality constraints are NP-hard \citep{natarajan1995sparse}, making them intractable for large-scale networked systems.
To address this challenge, various sparsity-promoting methods have been developed, including $\ell_1$-norm relaxations \citep{fardad2011sparsity,babazadeh2016sparsity}, gradient-based algorithms\citep{lian2017game,lian2018sparsity}, and alternating-projection approaches \citep{lin2013design,negi2020sparsity}.
These methods typically include three stages: initializing a feasible controller, identifying a desirable sparse structure, and improving closed-loop performance by solving the resulting fixed-structure control problem.


This work integrates the \gls{vndt} and an \gls{lq} performance objective with sparsity-promoting strategies (weighted $\ell_1$ norm and \gls{admm} with a cardinality penalty). 
Although existing sparsity-promoting methods mitigate the intractability of cardinality constraints, combining them with \gls{vndt} still results in an NP hard problem due to the presence of high-order matrix inequalities.
To resolve this challenge, we employ \gls{ico} \citep{warner2017iterative}, in which the high-order matrix inequalities are overbounded using the techniques in \citep{sebe2018sequential,ren2021successive}, and the resulting convex optimization problems are iteratively solved.
This approach yields a sparsely interconnected controller that is robust and applicable to networks of nonlinear and heterogeneous agents.

\section{Preliminaries}
\subsection{Notation}
The sets of real, natural numbers, and natural numbers up to $n$ are denoted by $\bbR$, $\bbN$, $\bbN_n$, respectively. 
The set of real $n{\times}m$ matrices is $\bbR^{n{\times} m}$, and the $(i,j)^\thh$ block or element of a matrix $\bA$ is denoted $\bA_{ij}$.
If $\bA_{ij}{\in}\bbR^{n_i{\times}m_i}$ and $\bA{\in}\bbR^{\sum_{i=1}^Nn_i{\times}\sum_{j=1}^Mm_j}$, then $\bA_{ij}$ is said to be a ``block'' of $\bA$, and $\bA$ is said to be in $\bbR^{N\times M}$ block-wise.
The set of $n{\times} n$ symmetric matrices is $\bbS^n$. 
The notation $\bA{\prec}0$ indicates that $\bA$ is negative-definite. 
For brevity, $\he(\bA){=}\bA{+}\bA^T$ and asterisks, $*$, denote duplicate blocks in symmetric matrices.
The set of square integrable functions is $\cL_{2}$. The Frobenius norm, $\cL_1$ norm, and $\cL_2$ norm are denoted by $\norm{\cdot}_F$, $\norm{\cdot}_1$ and $\norm{\cdot}_2$, respectively.  The truncation of a function $\textbf{y}(t)$ at $T$ is denoted by $\by_T(t)$, where $\by_T(t){=}\by(t)$ if $t{\leq}T$, and $\by_T(t){=}0$ otherwise. If $\|\by_T\|_2^2{=}{\langle}\by_T{,}\by_T{\rangle}{=}\int_0^{\infty}\by_T^T(t)\by_T(t)dt{<}\infty$ for all $T{\geq}0$, then $\by{\in}\cL_{2e}$, where $\cL_{2e}$ is the extended $\cL_2$ space.

\subsection{\texorpdfstring{$QSR$}{QSR}-Dissipativity of Large-Scale Systems}
This section reviews the $QSR$-dissipativity and its application to stability analysis of networks. 
\begin{definition} [$QSR$-Dissipativity, \citep{vidyasagar1981input}] \label{def:dissipativity}
    Let $\bQ\in\bbS^l$, $\bR\in\bbS^m$, and $\bS\in\bbR^{l\times m}$. The operator $\scrG:\cL_{2e}^m\mapsto\cL_{2e}^l$ is $QSR$-dissipative if there exists $\beta\in\bbR$ such that for all $\bu\in\cL_{2e}^m$ and $T$.
    \begin{align} \label{eq:dissipativity}
        \int_0^T\quadsupplyinput{\bQ}{\bS}{\bR}{\scrG(\bu(t))}{\bu(t)}dt\geq0
    \end{align}
\end{definition}
We identify $QSR$-dissipativity of \gls{lti} systems using the following lemma. Note that the stability assumption in \citep{gupta1996robust} was only needed for necessity. Hence, it provides sufficient conditions for dissipativity of both stable and unstable systems.
\begin{lemma} [Dissipativity Lemma, \citep{gupta1996robust}] \label{lem:kyp_lemma}
    An \gls{lti} system with minimal state-space realization $(\bA,\bB,\bC,\bD)$ is $QSR$-dissipative if there exist a matrix $\bP\succ0$ and matrices $\bQ$, $\bS$, and $\bR$ such that
    \begin{align} \label{eq:kyp_lemma}
        \begin{bmatrix}
            \he(\bP\bA){-}\bC^T\bQ\bC & \bP\bB{-}\bC^T\bS{-}\bC^T\bQ\bD \\
            \bB^T\bP{-}\bS^T\bC{-}\bD^T\bQ\bC & {-}\bR{-}\he(\bS^T\bD){-}\bD^T\bQ\bD
        \end{bmatrix}{\preceq}0
    \end{align}
\end{lemma}

Dissipativity is closely related to the \gls{io}-stability of the system, defined below.
\begin{definition} [\gls{io} or $\cL_2$-stability, \citep{lozano2013dissipative}]
    An operator $\scrG:\cX_{e}^m\mapsto\cX_{e}^l$ is \gls{io}-stable, if for any $\bu\in\cX^m$ and all $\bx_0$ where $\cX$ is any semi-inner product space and $\cX_e$ is its extension, there exists a constant $\kappa>0$ and a function $\beta(\bx_0)$ such that
    \begin{align}
        \norm{(\scrG(\bu))_T}_\cX\leq\kappa\norm{\bu_T}_\cX+\beta(\bx_0)
    \end{align}
    where $\norm{\cdot}_\cX$ is the induced norm of the innerproduct space. If the space $\cX$ is $\cL_2$, then \gls{io} stability is called $\cL_2$ stability.
\end{definition}
The properties of $(\bQ,\bS,\bR)$ of dissipative systems are closely related to the $\cL_2$-stability through \autoref{thm:stability_thm}.
\begin{theorem} \label{thm:stability_thm}
    The operator is $\cL_2$ stable if and only if it is $QSR$-dissipative with $\bQ\prec0$.
\end{theorem}

Dissipativity serves as a powerful framework for $\cL_2$-stability analysis of large-scale, multi-agent systems.
\begin{theorem} [\gls{vndt}, \citep{moylan2003stability}] \label{thm:vndt}
    Consider $N$ dissipative operators, $\scrG_i{:}\cL_{2e}^{m_i}{\mapsto}\cL_{2e}^{l_i}$ with parameters $(\bQ_i\bS_i\bR_i)$, interconnected by the operators, $\scrH_{ij}:\cL_{2e}^{l_j}{\mapsto}\cL_{2e}^{m_i}$, mapping the output of $j^\thh$ operator to the input of $i^\thh$ operator and represented by matrices $\bH_{ij}{\in}\bbR^{m_i\times l_j}$. Thereby, the large-scale interconnected system is expressed by 
    \begin{align} \label{eq:interconnected_systems}
        \by_i=\scrG\be_i,\quad\by=\scrG\bu,\quad\be=\bu+\bH\by,
    \end{align}
    where $\bu{=}\col(\bu_i)_{i\in\bbN_N}$, $\by{=}\col(\by_i)_{i\in\bbN_N}$, $\be{=}\col(\be_i)_{i\in\bbN_N}$, and $\scrG{=}\diag(\scrG_i)_{i\in\bbN_N}$. Then, $\scrG{:}\cL_{2e}^m{\mapsto}\cL_{2e}^l$ is $\cL_2$ stable if $\hatbQ{\prec}0$, where
    \begin{align} \label{eq:network_dissipativity_thm}
        \hatbQ=\bQ+\bS\bH+\bH^T\bS^T+\bH^T\bR\bH
    \end{align}
    with $\bQ{=}\diag(\bQ_i)_{i\in\bbN_N}$, and $\bS$ and $\bR$ defined analogously.
\end{theorem}


\subsection{Convex Overbounding} \label{subChap:convex_overbounding}
Optimal controller synthesis minimizes an objective function subject to matrix inequality constraints that are typically \glspl{bmi}, expressed as%
\begin{align} \label{eq:bmi}
    \bQ+\he(\bX\bN\bY)\prec0,
\end{align}
where $\bN{\in}\bbR^{p{\times}q}$ is a given matrix, and $\bQ{\in}\bbS^n$, $\bX{\in}\bbR^{n{\times}p}$. and $\bY{\in}\bbR^{q{\times}n}$ are design variables.
Since \autoref{eq:bmi} is generally nonconvex, the associated optimal control problem becomes NP-hard.
Conservative convexifications of \autoref{eq:bmi} can be derived using the following results.
\begin{theorem} [\citep{sebe2018sequential},\citep{ren2021successive}] \label{thm:overbounding}
    Consider the matrices $\bQ{\in}\bbS^n$, $\bN{\in}\bbR^{p\times q}$, $\bX{\in}\bbR^{n\times p}$. and $\bY{\in}\bbR^{q\times n}$, where $\bQ$, $\bX$, and $\bY$ are design variables.
    The \gls{bmi} condition $\bQ{+}\he(\bX\bN\bY){\prec}0$ is implied by either
    \begin{align} \label{eq:overbounding_sebe}
        \begin{bmatrix}
            \bQ & \bX\bN{+}\bY^T\bG^T \\
            \bN^T\bX^T{+}\bG\bY & -\he(\bG)
        \end{bmatrix}{\prec}0
    \end{align}
    for any $\bG{\in}\bbR^{q{\times}p}$ satisfying $\he(\bG){\succ}0$, or
    \begin{align} \label{eq:overbounding_ren}
        \begin{bmatrix}
            \bQ & * & * & \bzero \\
            \bN^T\bX^T{+}\bG^0\bY & -\he(\bG^0{+}\dbG) & \bzero & * \\
            \bY & \bzero & -\bF & 0 \\
            \bzero & \bF^0\dbG^T & \bzero & -2\bF^0{+}\bF
        \end{bmatrix}{\prec}0
    \end{align}
    for given $\bG^0{\in}\bbR^{q{\times}p}$ such that $\he(\bG^0){\prec}0$ and $\bF^0{\in}\bbS^q$, where $\dbG$ and $\bF$ are additional design variables.
\end{theorem}

Directly applying \autoref{thm:overbounding} may introduce excessive conservatism, making it difficult to obtain a feasible solution to \autoref{eq:bmi}.
Fortunately, its conservatism can be significantly reduced by linearizing around a feasible point, $\bX^0$ and $\bY^0$ satisfying $\bQ{+}\he(\bX^0\bN\bY^0){\prec}0$, if one is known. Specifically, \autoref{thm:overbounding} can be reformulated by substituting $\bQ$, $\bX$, and $\bY$ with $\bQ{+}\he(\bX^0\bN\bY^0{+}\dbX\bN\bY^0{+}\bX^0\bN\dbY)$, $\dbX$, and $\dbY$, and solve \autoref{eq:overbounding_sebe} or \autoref{eq:overbounding_ren} with new design variables $\bQ$, $\dbX$, and $\dbY$.
The resulting problem is always feasible since $\dbX{=}\bzero$ and $\dbY{=}\bzero$ yields 
the initial feasible point $(\bX^0,\bY^0)$.

\section{Sparsity-Promoting Dissipativity-Augmented Control}
\begin{figure}
    \captionsetup[sub]{aboveskip=2pt, belowskip=0pt}
    \centering
    \subfloat[Networked system]{\includegraphics[width=0.23\textwidth]{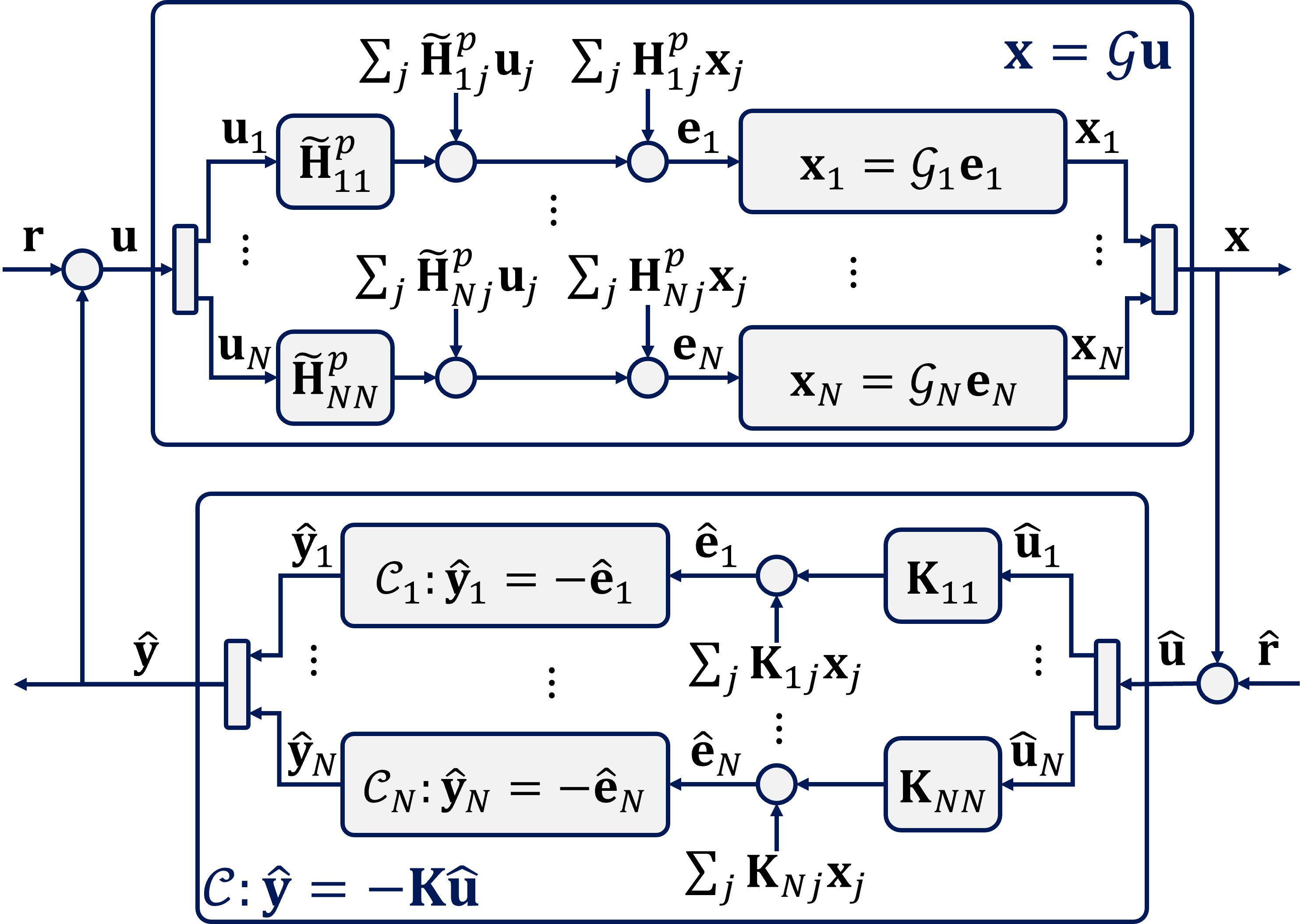}\label{fig1:decomposition}}
    \subfloat[Decomposed system]{\includegraphics[width=0.25\textwidth]{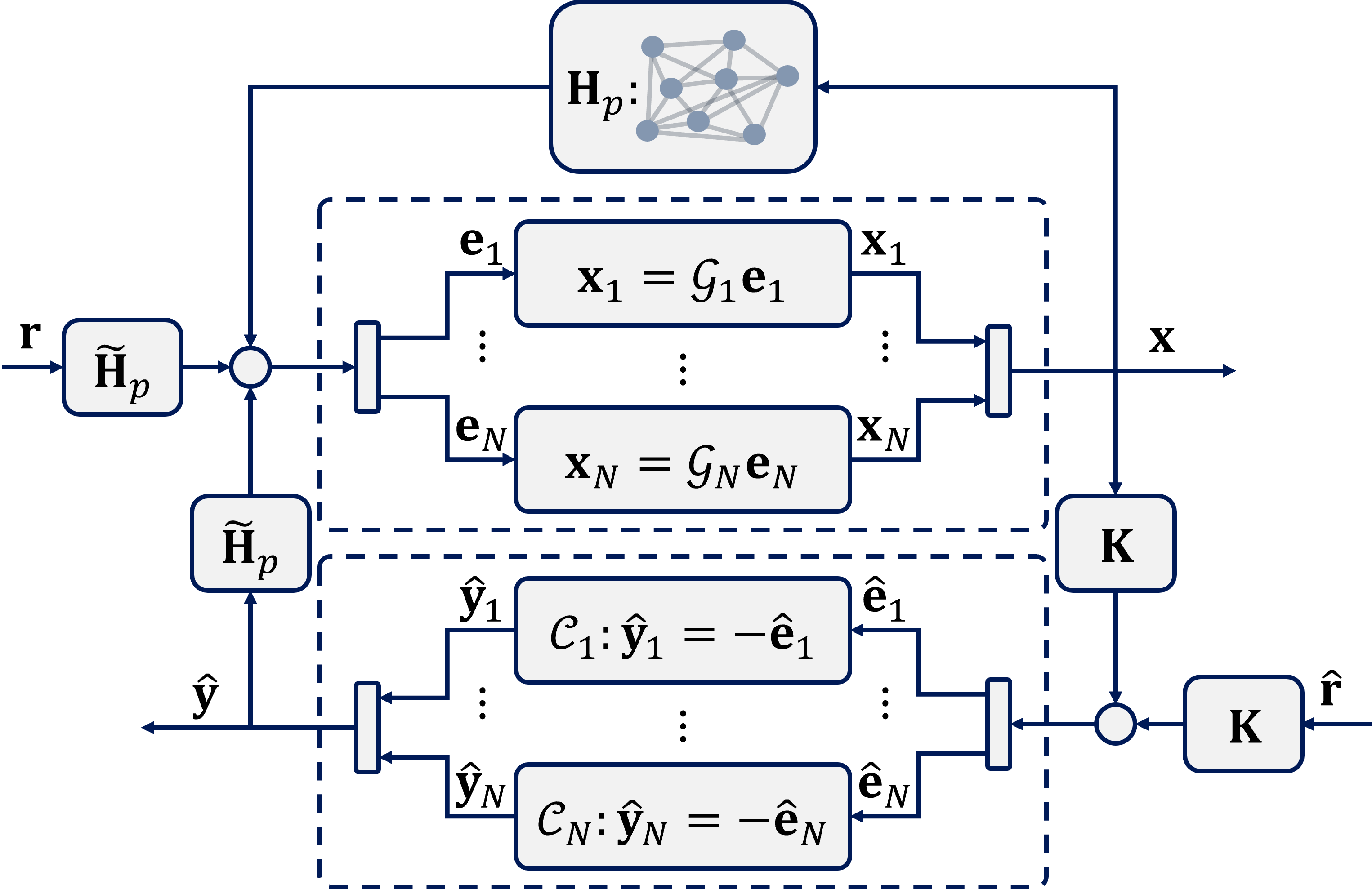}\label{fig1:equivalent}}
    \vspace{-8pt}
    \caption{Two representations of a closed-loop system.}
    \label{fig1:decomposition_K}
\end{figure}

Consider a plant $\scrG{:}\bu{\to}\bx$, where $\bu$ and $\bx$ are the input and state. 
The plant is controlled by a full-state feedback controller $\scrC{:}\hatbu{\to}\hatby$, expressed as $\hatby{=}{-}\bK\hatbu$. Disturbances may influence both the plant and controller inputs, denoted by $\br$ and $\hatbr$, respectively, so that $\bu{=}\br{+}\hatby$ and $\hatbu{=}\hatbr{+}\bx$, respectively. 
Suppose that the feedback interconnection of $\cG$ and $\cC$ can be decomposed into: 
$N$ agents, $\scrG_i{:}\be_i{\to}\bx_i$, where $\be_i$ and $\bx_i$ are the input and state of the $i^\thh$ agent; their local controllers, $\scrC_i{:}\hatbe_i{\to}\hatby_i$, expressed as $\hatby_i{=}{-}\hatbe_i$; and their interconnection gains.
The interconnection between the agents and their controller is given by
\begin{align*}
    \be_i&{=}\hspace{-10pt}\sum_{j\in\bbN_N,i\neq j}\hspace{-10pt}\bH_{ij}^p\bx_j{+}\hspace{-5pt}\sum_{j\in\bbN_N}\hspace{-3pt}\tilbH_{ij}^p\bu_j,\quad
    \hatbe_i{=}\hspace{-5pt}\sum_{j\in\bbN_N}\hspace{-3pt}\bK_{ij}\hatbu_{j}, \\
    \begin{bmatrix}
        \be \\ \hatbe
    \end{bmatrix}
    &{=}\begin{bmatrix}
        \tilbH_p\br \\ \bK\hatbr
    \end{bmatrix}+\begin{bmatrix}
        \bH_p & \tilbH_p \\ \bK & \bzero
    \end{bmatrix}
    \begin{bmatrix}
        \bx \\ \hatby
    \end{bmatrix}
    {=}\begin{bmatrix}
        \tilbH_p\br \\ \bK\hatbr
    \end{bmatrix}+\bH\begin{bmatrix}
        \bx \\ \hatby
    \end{bmatrix},
\end{align*}
where $\be=\col(\be_i)_{i\in\bbN_N}$, and $\hatbe$, $\bx$, and $\hatby$ are defined analogously. $\bH_p$ characterizes the interconnection structure among agents, $\tilbH_p$ represents the interconnection structure from the input of the plant to the input of each agent, and $\bH_{ij}^p$ and $\tilbH_{ij}^p$ are the corresponding block components.
\autoref{fig1:decomposition_K} depicts this system decomposition.
The objective is to synthesize a sparse controller $\bK$ that stabilizes the overall interconnected system in the presence of disturbances $\br$ and $\hatbr$, while minimizing a standard \gls{lq} criterion.

\subsection{Dual-Model Synthesis}
Stability is ensured by enforcing $QSR$-dissipativity properties on each agent and its local controller, $\cG_i$ and $\cC_i$, and by applying \gls{vndt}, \autoref{thm:vndt}. 
Although the external inputs to the decomposed network are $\tilbH_p\br$ and $\bK\hatbr$, \autoref{thm:vndt} remains applicable because signals remain in $\cL_2$ after premultiplication by a constant gain.

The \gls{lq} performance is evaluated based on the closed-loop dynamics of the linearized plant, $\cG^{LTI}$, and controller, 
\begin{align*}
    \cG^{LTI}:\dbx=\bA\bx+\bB\bu, \quad
    \cC:\bu=-\bK\bx,
\end{align*}
where $\bA$ and $\bB$ denote the linearized plant matrices.
Synthesizing $\bK$ matrix that minimizes \gls{lq} cost of the closed-loop system is equivalent to solving
\begin{align*}
    \arg\min_{\scriptstyle\substack{\bK,\bP}}\;\;&\tr(\bP) \\[-4pt]
    \st\;\;&\he(\bP(\bA{-}\bB\bK)){+}\bQ_{lq}{+}\bK^T\bR_{lq}\bK{\preceq}0,\;\bP{\succ}0
\end{align*}
where $\bQ_{lq}\succ0$ and $\bR_{lq}\succ0$ are state and control weight of the \gls{lq} criterion \citep{khalil1996robust}.
To promote sparsity in $\bK$, a penalty of $g(\bK)$ is added to the objective. 
Typical sparsity-promoting penalty functions include the $\ell_1$ norm, weighted $\ell_1$ norm, sum-of-logs, and the cardinality function of $\bK$ \citep{lin2013design}.

In conclusion, the sparsity-promoting dissipativity-augmented controller is obtained by solving 
\begin{subequations} \label{eq:main_problem}
\begin{align}
    \arg\min_{\scriptstyle\substack{\bK,\bP}}\;&\tr(\bP){+}\gamma g(\bK) \label{eq:objective}\\[-4pt]
    \st\;&\he(\bP(\bA{-}\bB\bK)){+}\bQ_{lq}{+}\bK^T\bR_{lq}\bK{\preceq}0,\;\bP{\succ}0, \label{eq:lyapunov}\\
    &\int_0^T\quadsupplyinput{\bQ_i}{\bS_i}{\bR_i}{\by_i}{\be_i}dt{\geq}0,\;\forall i{\in}\bbN_N, \label{eq:diss_plant}\\
    &{-}\hatbR_i{+}\he(\hatbS_i){-}\hatbQ_i{\preceq}0,\;\forall i{\in}\bbN_N, \label{eq:diss_ctrl}\\
    &\bQ{+}\he(\bS\bH){+}\bH^T\bR\bH{\prec}0. \label{eq:diss_sys}
\end{align}
\end{subequations}
where $\gamma$ denotes a penalty parameter, $(\bQ_i,\bS_i,\bR_i)$ and $(\hatbQ_i,\hatbS_i,\hatbR_i)$ respectively represent $QSR$ parameters of $i^\thh$ agent and its local controller, and $\bQ=\diag(\diag(\bQ_i)_{i\in\bbN_N},\diag(\hatbQ_i)_{i\in\bbN_N})$,  while $\bS$ and $\bR$ are defined analogously.
However, \autoref{eq:main_problem} is non-convex due to \autoref{eq:lyapunov}, \autoref{eq:diss_plant}, and \autoref{eq:diss_sys}.
These constraints are tightened to \glspl{lmi} in \autoref{Chap:main_results}.

\section{Main Results} \label{Chap:main_results}
This section proposes strategies to solve \autoref{eq:main_problem}. 
\autoref{eq:diss_plant} is reformulated to \glspl{lmi} by applying the expansions from \autoref{lem:kyp_lemma}, based on the structures of agents. 
\autoref{eq:lyapunov}, and \autoref{eq:diss_sys} are reconstructed using the convex overbounding techniques introduced in \autoref{subChap:convex_overbounding}.
\autoref{subChap:Sparsity_Promotion} provides the approach to compute a feasible sparse controller that improves the \gls{lq} performance criterion, based on \gls{admm}.


\subsection{Agent Dissipativity} \label{subChap:Agent_Dissipativity}
\autoref{def:dissipativity} cannot be directly applied to solve \autoref{eq:main_problem} using existing \gls{sdp}-based optimization solvers. 
Fortunately, many variations of \autoref{def:dissipativity} result in tractable \glspl{lmi} that enforce the $QSR$-dissipativity of certain nonlinear systems.
For agents $\cG_i$ modeled as \gls{lti} systems, \autoref{lem:kyp_lemma} can be applied to impose the dissipativity constraint.
If $\cG_i$ is an \gls{lti} system with polytopic uncertainty in its dynamics, \citep[Equation 18]{walsh2019interior} can be used to verify the $QSR$-dissipativity of the agent.
If $\cG_i$ is an \gls{lti} system with input, output or state delays, \citep[Theorem 3.1]{bridgeman2016conic} provides \glspl{lmi} to determine dissipativity parameters of the agent. Gain of control affine systems can be determined from \glspl{lmi} following \cite{Strong2024IterativeGain}.

In practice, dissipativity of nonlinear agents is often established a priori through ad hoc analysis, or prior knowledge. For instance, Eulter-Lagrange systems are passive, meaning $\bQ_i^p=\bzero$, $\bS_i^p=\frac{1}{2}\bI$, and $\bR_i^p=\bzero$ \citep[Chapter 6]{lozano2013dissipative}.
In addition, data-driven methods such as \citep{romer2017determining} can be employed to estimate $(\bQ_i^p,\bS_i^p,\bR_i^p)$ for the agents. Then the predefined dissipativity parameters $(\bQ_i^p,\bS_i^p,\bR_i^p)$ can be scaled by a design variable $\lambda{\geq}0$, so that $\bQ_i{=}\lambda\bQ_i^p$, $\bS_i{=}\lambda\bS_i^p$, and $\bR_i{=}\lambda\bR_i^p$ to search over equivalent characterizations.

\subsection{Convex Overbounding of \texorpdfstring{\gls{bmi}}{BMI} Constraints} \label{subChap:Convex_Overbounding_of_BMI_Constraints}
Corollaries \ref{cor:lmi_lyapunov} and \ref{cor:lmi_vndt} apply \autoref{thm:overbounding} to yield \glspl{lmi} implying \autoref{eq:lyapunov}, and \autoref{eq:diss_sys}.
\begin{corollary} \label{cor:lmi_lyapunov}
    Given $\bK^0$ and $\bP^0$, suppose there exist $\dbK$, $\dbP$, $\bG^0$ and $\bF^0{\succeq}0$ satisfying $\he(\bG^0){\succeq}0$ and
    \begin{align} \label{eq:lmi_lyapunov}
        \hspace{-9pt}{\setlength{\arraycolsep}{0pt}
        \begin{bmatrix}
            \bL{+}\bQ_{lq} & * & * & * & \bzero \\
            \bK^0{+}\dbK & {-}\bR_{lq}^{{-}1} & \bzero & \bzero & \bzero \\
            \hspace{-3pt}{(}\bB\dbP{)}^{\hspace{-1pt}T}\hspace{-3pt}{-}\bG^0\dbK & \bzero & {-}\he(\bG^0{+}\dbG) & \bzero & * \\
            {-}\dbK & \bzero & \bzero & -\bF & \bzero \\
            \bzero & \bzero & \bF^0\dbG^T & \bzero & {-}2\bF^0{+}\bF
        \end{bmatrix}\hspace{-3pt}{\preceq}0
        }{,}\hspace{-6pt}
    \end{align}
    where $\bL{=}\he\set{\hspace{-1.5pt}(\bP^0{+}\dbP)(\bA{-}\bB\bK^0){-}\bP^0\bB\dbK}\hspace{-1.5pt}$. Then $\bK^0+\dbK$ and $\bP^0{+}\dbP$ are feasible points of \autoref{eq:lyapunov}. Moreover, if $\bK^0$ and $\bP^0$ are feasible for \autoref{eq:lyapunov}, \autoref{eq:lmi_lyapunov} is always feasible.
\end{corollary}
\begin{pf}
    The proof follows by applying Schur complement and the overbounding condition in \autoref{eq:overbounding_ren} of \autoref{thm:overbounding}. Since $\bR_{lq}{\succ}0$, applying Schur complement to \autoref{eq:lyapunov} and substituting $\bK$ and $\bP$ with $\bK^0{+}\dbK$ and $\bP^0{+}\dbP$ yield
    \begin{align*}
        \begin{bmatrix}
            \bL{+}\bQ_{lq} & * \\
            \bK^0{+}\dbK & {-}\bR_{lq}^{-1}
        \end{bmatrix}
        {+}\he\brackets{
            \begin{bmatrix}
                \dbP\bB \\ \bzero
            \end{bmatrix}
            \begin{bmatrix}
                {-}\dbK & \bzero
            \end{bmatrix}
        }{\preceq}0
    \end{align*}
    which is equivalent to \autoref{eq:lyapunov}.
    Next, applying \autoref{eq:overbounding_ren} in \autoref{thm:overbounding} with $\bX{=}[(\dbP\bB)^T\;\bzero]^T$, $\bY{=}[-\dbK\;\bzero]$, and $\bN{=}\bI$ yields \autoref{eq:lmi_lyapunov}.
    The feasibility of $\bK^0$ and $\bP^0$ implies that $\dbK{=}\bzero$, $\dbP{=}\bzero$, $\dbG{=}\bzero$, and $\bF{=}\bI$ are a feasible solution of \autoref{eq:lmi_lyapunov}.
    $\hfill\blacksquare$
\end{pf}
\begin{corollary} \label{cor:lmi_vndt}
    Given $\bQ^0$, $\bS^0$, $\bR^0$, and $\bH^0$, suppose there exists $\dbQ$, $\dbS$, $\dbR$, and $\dbH$ such that
    \begin{align} \label{eq:lmi_vndt}
        \begin{bmatrix}
            \bM & * & * & * \\
            \dbS^T{+}\dbH & {-}2\bI & \bzero & \bzero \\
            \frac{1}{2}\bR^0\dbH{+}\dbH & \bzero & {-}2\bI & * \\
            \dbR\bH^0{+}\dbH & \bzero & \frac{1}{2}\dbR & {-}2\bI
        \end{bmatrix}{\prec}0,
    \end{align}
    where $\bM{=}\bQ^0\hspace{-1pt}{+}\dbQ\hspace{-1pt}{+}\hspace{-1pt}\he(\hspace{-1pt}\bS^0\bH^0{+}\dbS\bH^0{+}\bS^0\dbH\hspace{-1pt}){+}\bH^{0T}\hspace{-1.5pt}\bR^0\bH^0\hspace{-5pt}+\bH^{0T}\hspace{-1pt}\dbR\bH^0\hspace{-1.5pt}{+}\he(\hspace{-1pt}\dbH^T\bR^0\bH^0\hspace{-1pt})$.
    Then $\bQ^0\hspace{-2pt}{+}\dbQ$, $\bS^0\hspace{-2pt}{+}\dbS$, $\bR^0\hspace{-2pt}{+}\dbR$, and $\bH^0\hspace{-2pt}{+}\dbH$ are a feasible solution of \autoref{eq:diss_sys}.
    Moreover, if $\bQ^0$, $\bS^0$, $\bR^0$, and $\bH^0$ are feasible for \autoref{eq:diss_sys}, \autoref{eq:lmi_vndt} is always feasible.
\end{corollary}
\begin{pf}
    Since \autoref{eq:diss_sys} is a trilinear matrix inequality
    , \autoref{eq:lmi_vndt} is obtained by applying \autoref{eq:overbounding_sebe} of \autoref{thm:overbounding} twice to \autoref{eq:diss_sys}.
    First, using $\bS{=}\bS^0{+}\dbS$ and $\bH{=}\bH^0{+}\dbH$ allows \autoref{eq:diss_sys} to be reformulated as
    \begin{align*}
        &\bQ{+}\he(\bS^0\bH^0{+}\dbS\bH^0{+}\bS^0\dbH){+}\bH^{0T}\bR\bH^0{+}\he(\dbH^T\bR\bH^0)\\
        &\quad{+}\he\brackets{\begin{bmatrix}
            \dbS & \frac{1}{2}\dbH^T\bR
        \end{bmatrix}
        \begin{bmatrix}
            \dbH \\ \dbH
        \end{bmatrix}
        }{\prec}0.
    \end{align*}
    Applying \autoref{eq:overbounding_sebe} of \autoref{thm:overbounding} with $\bX=[\dbS\;\frac{1}{2}\dbH^T\bR]$, $\bY{=}[\dbH^T\;\dbH^T]^T$, $\bN{=}\bI$ and $\bG{=}\bI$, and replacing $\bQ$ and $\bR$ with $\bQ^0{+}\dbQ$ and $\bR^0{+}\dbR$, respectively, lead to
    \begin{align*}
        \begin{bmatrix}
            \bM & * & * \\
            \dbS^T{+}\dbH & -2\bI & \bzero \\
            \frac{1}{2}\bR^0\dbH{+}\dbH & \bzero & -2\bI
        \end{bmatrix}
        {+}\he\brackets{
            \begin{bmatrix}
                \bH^{0T}\dbR \\ \bzero \\ \frac{1}{2}\dbR
            \end{bmatrix}
            \begin{bmatrix}
                \dbH & \bzero & \bzero
            \end{bmatrix}
        }{\prec}0
    \end{align*}
    Finally, applying \autoref{eq:overbounding_sebe} of \autoref{thm:overbounding} once more with $\bX{=}[\dbR^T\bH^0\;\bzero\;\frac{1}{2}\dbR^T]^T$, $\bY{=}[\dbH\;\bzero\;\bzero]$, $\bN{=}\bI$, and $\bG{=}\bI$ results in \autoref{eq:lmi_vndt}.    
    The feasibility of $\bQ^0$, $\bS^0$, $\bR^0$ and $\bH^0$ implies that $\dbQ{=}\bzero$, $\dbS{=}\bzero$, $\dbR{=}\bzero$, and $\bH{=}\bzero$ are a feasible solution of \autoref{eq:lmi_vndt}.
    $\hfill\blacksquare$
\end{pf}
For the application of \autoref{cor:lmi_vndt} to \autoref{eq:diss_sys}, the design variables are defined as
\begin{align*}
    \dbH=\begin{bmatrix}
        \bzero & \bzero \\ \dbK & \bzero
    \end{bmatrix},\;\;
    {\setlength{\arraycolsep}{-5pt}
    \dbQ{=}\begin{bmatrix}
        \diag(\dbQ_i)_{i\in\bbN_N} & \bzero \\
        \bzero & \diag(\hatdbQ_i)_{i\in\bbN_N}
    \end{bmatrix}},
\end{align*}
and $\dbS$ and $\dbR$ are defined like $\dbQ$.
Applying \autoref{cor:lmi_lyapunov} and \autoref{cor:lmi_vndt}, we have the problem 
\begin{subequations} \label{eq:main_problem_convex_central}
    \begin{align}
        \arg\!\!\min_{\substack{
        \scriptscriptstyle \dbK,\dbP,\dbQ,\\
        \scriptscriptstyle \dbS,\dbR,\dbG,\bF
    }}\;&\tr(\bP^0+\dbP) \label{eq:objective_convex_central}\\[-4pt]
        \st\;\;\;&\bN(\dbQ,\dbS,\dbR,\dbK,\dbP,\dbG,\bF)\preceq0, \label{eq:constraints_convex_central}
    \end{align}
\end{subequations}
to construct the centralized controller for given initial feasible point, where $\bN(\dbQ,\dbS,\dbR,\dbK,\dbP,\dbG,\bF)$ composes: \autoref{eq:lmi_lyapunov} to ensure that $\tr(\bP^0+\dbP)$ bounds the \gls{lq} norm above; agent dissipativity requirements encoded as \glspl{lmi}, as discussed in \autoref{subChap:Agent_Dissipativity}; \autoref{eq:diss_ctrl} to impose controller dissipativity; and \autoref{eq:lmi_vndt} to ensure that the network is stable via \gls{vndt}.

\subsection{Initialization} \label{subChap:Initialization}
Initial feasible points of \autoref{eq:main_problem} or \autoref{eq:main_problem_convex_central}, denoted by $\bP^0$, $\bK^0$, $\bQ^0$, $\bS^0$, $\bR^0$, $\hatbQ^0$, $\hatbS^0$, $\hatbR^0$, $\bG^0$, and $\bF^0$, are required to solve \autoref{eq:main_problem_convex_central}. 
Heuristically, localized controllers $\bK_i{\in}\bbR^{}$ for $i{\in}\bbN_N$ can be initialized based on classical control methods, such as PD control, and a feasible point of \autoref{eq:main_problem} can then be obtained with these fixed local controllers.
If the plant agents are stable, the strategy in \citep[Section 6]{locicero2025dissipativity} can be applied.
If the constraints are not satisfied for a chosen localized controller, the feasibility check is repeated with increased controller gains until a feasible point is found.
Once a feasible set of local controller is obtained, solving \autoref{eq:main_problem_convex_central} results in the initial feasible point corresponding to a centralized initial controller.
Alternatively, the iterative relaxation strategy proposed in \citep{warner2017iterative} can also be employed to obtain an initial feasible solution.
For $\bG^0$ and $\bF^0$, any matrices satisfying conditions in \autoref{cor:lmi_lyapunov} can be used, and $\bI$ is commonly used for the initialization.


\subsection{Sparsity Promotion} \label{subChap:Sparsity_Promotion}
Once the initial centralized feasible points are obtained, the optimal sparse structure can be determined by 
\begin{subequations} \label{eq:main_problem_convex_sparse}
    \begin{align}
        \arg\!\!\min_{\substack{
        \scriptscriptstyle \dbK,\dbP,\dbQ,\\
        \scriptscriptstyle \dbS,\dbR,\dbG,\bF
    }}\;&\tr(\bP^0+\dbP){+}\gamma g(\bK^0+\dbK) \label{eq:objective_convex}\\[-4pt]
        \st\;\;\;&\bN(\dbQ,\dbS,\dbR,\dbK,\dbP,\dbG,\bF)\preceq0, \label{eq:constraints_convex}
    \end{align}
\end{subequations}
where a penalty function $g(\bK^0{+}\dbK)$ is introduced to promote sparsity in the feedback gain. 
Several forms can be considered as the penalty function. 
In this work, the cardinality function and weighted $\ell_1$ norm are employed, since the weighted $\ell_1$ norm provides an appropriate convex approximation of the cardinality penalty \citep{lin2013design}.

The weighted $\ell_1$ norm is defined as 
\begin{align} \label{eq:weighted_ell1}
    g(\bK^0+\dbK)&=\hspace{-5pt}\sum_{i,j\in\bbN_N}\hspace{-5pt}W_{ij}\norm{\bK_{ij}^0+\dbK_{ij}}_F
\end{align}
where $W_{ij}$ are non-negative weights. 
If $W_{ij}$ are inversely proportional to $\norm{\bK_{ij}^0{+}\dbK_{ij}}_F$, the weighted $\ell_1$ norm becomes equivalent to the cardinality penalty.
However, $W_{ij}$ cannot depend on the unknown variables \citep{lin2013design}.
Instead, the structure of the initial feasible controller $\bK^0$ can be used to define $W_{ij}{=}\min\{ \norm{\bK^0_{ij}}_F^{-1}{,}\ \epsilon_l^{-1}\}$.
When this weighted $\ell_1$ norm is used, the penalty function becomes convex, so \autoref{eq:main_problem_convex_sparse} can be solved numerically.

\cite{lin2013design} proposed to solve the problem with a cardinality penalty using \gls{admm}.
Although \gls{admm} does not guarantee convergence for nonconvex objectives, extensive computational experience has shown that it performs well in practice when the augmented Lagrangian parameter is sufficiently large \citep{boyd2011distributed}. We therefore adopt the strategy of \cite{lin2013design} for \autoref{eq:main_problem_convex_sparse}, using 
\begin{align} \label{eq:cardinality_penalty}
    g(\bK^0+\dbK)=\hspace{-5pt}\sum_{i\in\bbN_N}\hspace{-2pt}\sum_{j\in\bbN_N}\hspace{-4pt}\card\brackets{\norm{\bK^0_{ij}+\dbK_{ij}}_F}.
\end{align}
By introducing the indicator function $I_\bN$ and an auxiliary variable $\bZ$, where $I_\bN$ takes $0$ if $\bN\preceq0$ and $+\infty$ otherwise, \autoref{eq:main_problem_convex_sparse} is equivalently written as
\begin{align*}
    \arg\!\!\!\!\!\!\min_{\substack{
        \scriptscriptstyle \dbK,\dbP,\bZ,\dbG\\
        \scriptscriptstyle \dbQ,\dbS,\dbR,\bF
    }}\;&J(\dbP){+}I_\bN(\dbQ,\dbS,\dbR,\dbK,\dbP,\dbG,\bF){+}\gamma g(\bZ) \\[-4pt]
    \st\;\;\;&\bK^0{+}\dbK{-}\bZ{=}\bzero,
\end{align*}
where $J(\dbP){=}\tr(\bP^0{+}\dbP)$.
\gls{admm} then solves
\begin{align*}
    \arg\!\!\!\!\!\!\min_{\substack{
        \scriptscriptstyle \dbK,\dbP,\bZ,\dbG\\
        \scriptscriptstyle \dbQ,\dbS,\dbR,\bF
    }}\;
    &\!\!\!\!J(\dbP){+}I_\bN(\dbQ,\dbS,\dbR,\dbK,\dbP,\dbG,\bF){+}\gamma g(\bZ) \\[-12pt]
    &{+}\tr\brackets{\bm\Lambda^T(\bK^0{+}\dbK{-}\bZ)}{+}\frac{\rho}{2}\norm{\bK^0{+}\dbK{-}\bZ}_F^2
\end{align*}
by iteratively solving
\begin{subequations} 
    \begin{align}
        \hspace{-5pt}\dbK^{k+1}&{=}\arg\!\!\!\!\!\min_{\substack{
            \scriptscriptstyle \dbK,\dbP,\dbQ,\\
            \scriptscriptstyle \dbS,\dbR,\dbG,\bF
        }}\!\!\!\!\!\tr(\bP^0{+}\dbP){+}\frac{\rho}{2}\hspace{-3pt}\norm{\bK^0{+}\dbK{-}\bZ^k{+}\bm\Lambda^k}_F^2 \nonumber\\[-4pt]
            &\qquad\quad\;\;\;\!\!\!\st\!\!\;\;\bN(\dbQ,\dbS,\dbR,\dbK,\dbP,\dbG,\bF){\preceq}0,  \label{eq:K_update}\\
        \bZ^{k+1}&{=}\arg\min_\bZ\gamma g(\bZ){+}\frac{\rho}{2}\hspace{-3pt}\norm{\bK^0{+}\dbK^{k+1}{-}\bZ{+}\bm\Lambda^k}_F^2{,}\hspace{-3pt} \label{eq:Z_update}\\
        \bm\Lambda^{k+1}&{=}\bm\Lambda^k{+}(\bK^0+\dbK^{k+1}{-}\bZ^{k+1}), \label{eq:Lambda_update}
    \end{align}
\end{subequations}
where $\bm\Lambda$ is the dual variable, $k$ is the iteration index, and $\rho{>}0$ is the augmented Lagrangian parameter.
If \autoref{eq:cardinality_penalty} is used as the penalty, the block component of the unique solution to \autoref{eq:Z_update}, $\bZ_{ij}^{k+1}$, is $\bV_{ij}$ if $\norm{\bV_{ij}}_F{>}\sqrt{2\gamma/\rho}$ and $\bzero$ otherwise,
where $\bV{=}\bK^0{+}\dbK^{k+1}{+}\bm\Lambda^k$ \citep{lin2013design}. The stopping criteria of \gls{admm} are $r_p{=}\frac{\norm{\bK^0{+}\dbK^{k}{-}\bZ^{k}}_F}{\norm{\bZ^{k}}_F}{\leq}\epsilon_p$ and $r_d{=}\frac{\norm{\bZ^{k}{-}\bZ^{k-1}}_F}{\norm{\bZ^{k}}_F}{\leq}\epsilon_d$.

After solving \autoref{eq:main_problem_convex_sparse}, the controller is updated to $\bK_1 {=} \bK^0 + \dbK^\star$, where $\dbK^\star$ is the optimal perturbation obtained from the solution.  
The matrices $\bQ^1$, $\bS^1$, $\bR^1$, $\hatbQ^1$, $\hatbS^1$, $\hatbR^1$, $\bY^1$, and $\bG^1$ are updated accordingly, and $\bF^1{=}\bF^\star$.
The subspace $\cK$ is then defined as the set of block matrices that share the same sparsity pattern as $\bK^1$.


\subsection{Structured \gls{ico}} \label{subChap:Structured_Iterative_Convex_Overbounding}
The process described in \autoref{subChap:Sparsity_Promotion} solves \autoref{eq:main_problem_convex_sparse}, which represents a conservative convexification of \autoref{eq:main_problem} about the initial feasible points.
This implies that the resulting solutions are not necessarily local optima of \autoref{eq:main_problem}.
Intuitively, the closed-loop performance can be further improved by convexifying about the new solution points and re-solving \autoref{eq:main_problem_convex_sparse}.
However, such an approach does not preserve the convergence guarantees of \gls{ico}, since the cardinality operator is pointwise discontinuous at zero, and no approximation of the cardinality can simultaneously be an overbound, convex, and nonconservative at zero \citep{locicero2022sparsity}.
Therefore, instead of iteratively solving \autoref{eq:main_problem_convex_sparse}, we consider
\begin{subequations} \label{eq:main_problem_structured}
    \begin{align}
        \arg\!\!\min_{\substack{
        \scriptscriptstyle \dbK,\dbP,\dbQ,\\
        \scriptscriptstyle \dbS,\dbR,\dbG,\bF
    }}\;&\tr(\bP^k+\dbP) \label{eq:objective_structured}\\[-4pt]
        \st\;\;\;&\bN(\dbQ,\dbS,\dbR,\dbK,\dbP,\dbG,\bF)\preceq0, \label{eq:constraints_structured} \\
        &\bK^k+\dbK\in\cK,
    \end{align}
\end{subequations}
where $\cK$ is the subspace achieved from \autoref{subChap:Sparsity_Promotion}. 
This problem is used within the \gls{ico} framework to iteratively update the feasible points, yielding the optimal sparse controller $\bK^\star$.
The overall procedure of \autoref{Chap:main_results} for obtaining $\bK^\star$ is summarized in \autoref{alg:sparsity_promotion}.

\begin{algorithm}[tbp]
    \caption{Sparsity-Promoting Dissipativity-Augmented Controller Synthesis}\label{alg:sparsity_promotion}
    \begin{algorithmic}[1]
        \Require $g{,}\bY^0,\bK^0,\bQ^0,\bS^0,\bR^0,\hatbQ^0,\hatbS^0,\hatbR^0,\bG^0,\bF^0,\epsilon_p,\epsilon_d,\epsilon$
        \Ensure $\bK$
        \If {$g(\bK^0+\dbK)$ follows \autoref{eq:weighted_ell1}}
            \State Find $\dbK^\star$ by solving \autoref{eq:main_problem_convex_sparse}
        \ElsIf {$g(\bK^0+\dbK)$ follows \autoref{eq:cardinality_penalty}}
            \While {$r_p>\epsilon_p$ or $r_d>\epsilon_d$}
                \State Find $\dbK^{k+1}$ by solving \autoref{eq:K_update}
                \State Find $\bZ^{k+1}$ by solving \autoref{eq:Z_update}
                \State Find $\bm\Lambda^{k+1}$ by solving \autoref{eq:Lambda_update}
            \EndWhile
        \EndIf
        \State Set $\bK^1=\bK^0+\dbK^\star$, and update corresponding feasible points, $\bY^1$, $\bQ^1$, $\bS^1$, $\bR^1$, $\hatbQ^1$, $\hatbS^1$, $\hatbR^1$, and $\bG^1$ analogous to $\bK^1$, and set $\bF^1=\bF^\star$
        \State Define the structured subspace $\cK$ based on the sparsity pattern of $\bK^1$
        \While {$\frac{\tr(\bY^k)-\tr(\bY^{k+1})}{\tr(\bY_{k+1})}>\epsilon$}
            \State Solve \autoref{eq:main_problem_structured} using $\bK^k$, $\bY^k$, $\bQ^k$, $\bS^k$, $\bR^k$, $\hatbQ^k$, $\hatbS^k$, $\hatbR^k$, $\bG^k$, and $\bF^k$
            \State Set $\bK^{k+1}=\bK^k+\dbK^\star$, and other feasible points, $\bY^{k+1}$, $\bQ^{k+1}$, $\bS^{k+1}$, $\bR^{k+1}$, $\hatbQ^{k+1}$, $\hatbS^{k+1}$, $\hatbR^{k+1}$, and $\bG^{k+1}$ analogous to $\bK^1$, and set $\bF^{k+1}=\bF^\star$
        \EndWhile
    \end{algorithmic}
\end{algorithm}




\subsection{Convergence of \autoref{alg:sparsity_promotion}}
\autoref{alg:sparsity_promotion} combines two major stages, sparsity promotion, described in \autoref{subChap:Sparsity_Promotion}, and \gls{ico}, described in \autoref{subChap:Structured_Iterative_Convex_Overbounding}.
The \gls{ico} is always feasible and converges to a local optimum, since it retains at least one feasible point, which is the solution of the previous iteration, and the cost of the problem in each iteration is guaranteed to be no greater than that of this feasible point.
In contrast, the convergence behavior of the sparsity-promotion stage depends on the choice of the penalty function.
When the weighted $\ell_1$ norm is employed, the algorithm converges to the a specific structured controller because the problem is formulated as a \gls{sdp}, which can be solved using standard convex optimization solvers.
On the other hand, when the cardinality function is used, there is no convergence guarantee due to its inherent nonconvexity. 

\section{Numerical Examples}
In this example, sparse controllers for a networked system composed of $N{=}10$ agents with uncertain parameters are synthesized using \autoref{alg:sparsity_promotion}.
The $10$ agents are randomly distributed over a $10{\times}10$ grid, and their dynamics deviate randomly from their nominal models.
The nominal dynamics of each agent follow the model introduced in \citep{motee2008optimal}, expressed as
\begin{align*}
    \dbx_i=\hatbA_{ii}\bx_i+\sum_{j\neq i}e^{-\alpha(i-j)}\bx_j+\begin{bmatrix}
        0 \\ 1
    \end{bmatrix}\be_i
\end{align*}
where $\hatbA_{ii}$ are given by $\begin{bmatrix}
    1 & 1 \\ 1 & 2
\end{bmatrix}$ for $i{\in}\bbN_5$ and $\begin{bmatrix}
    -2 & 1 \\ 1 & -3
\end{bmatrix}$ otherwise, $\alpha{=}0.1823$, $\bx_i{\in}\bbR^2$, and $\be_i{\in}\bbR$.
Hence, the nominal dynamics of the first 5 agents are unstable, whereas those of the remaining five agents are stable.
From local dynamics, $\bH_p$ is the block matrix where $\bH^p_{ii}{=}\bzero_{2\times2}$ and $\bH^p_{ij}{=}e^{-\alpha(i-j)}\bI_{2}$, and $\tilbH_p{=}\diag([0\;1]^T)_{i\in\bbN_{10}}$.
The linearized system for studying \gls{lq} objective is defined using $\bA{=}\diag(\hatbA_{ii})_{i\in\bbN_{10}}{+}\bH$ and $\bB{=}\tilbH_p$. 

The actual dynamics $\bA_{ii}$ of each agent are uniformly distributed within $\pm30\%$ of their nominal values $\hatbA_{ii}$.
Accordingly, each agent's dynamics can be modeled as an \gls{lti} system with polytopic uncertainty, represented by a polyhedron with 16 vertices, where each vertex corresponds to the case in which one parameter of $\bA_{ii}$ attains either its maximum or minimum value within the uniform distribution.
Since some agents are unstable without control, \citep[Equation 18]{walsh2019interior} cannot be applied because it requires open-loop stability. 
Therefore, \autoref{lem:polytopic_diss} is used, with $\bA^j$ denoting the 8 vertices of the polytope and $\bB^j=\bI_2$ for all $j\in\bbN_8$.

\begin{lemma} \label{lem:polytopic_diss}
    The operator $\scrG{:}\cL_{2e}^m{\mapsto}\cL_{2e}^l$ defined by $\dbx(t)=\bA\bx(t){+}\bB\bu(t)$ and $\by(t){=}\bx(t)$, where $\bA{=}\sum_{j=1}^v\zeta_j\bA^j$ and $\bB{=}\sum_{j=1}^v\hspace{-2pt}\zeta_j\bB^j$ with $\sum_{j=1}^v\hspace{-2pt}\zeta_j{=}1$, is $QSR$-dissipative if there exist a matrix $\bP{\succ}0$ and matrices $\bQ$, $\bS$, and $\bR$ such that
    \begin{align} \label{eq:polytopic_diss}
    \begin{bmatrix}
        \he(\bP\bA^j)-\bQ & \bP\bB^j-\bS \\ (\bP\bB^j)^T-\bS^T & -\bR
    \end{bmatrix}\preceq0,\quad\forall j\in\bbN_v.
    \end{align}
\end{lemma}
\begin{pf}
    The proof proceeds by verifying \autoref{lem:kyp_lemma}. The operator is $QSR$ dissipative if \autoref{lem:kyp_lemma} holds for $(\bA,\bB,\bI,\bzero)$.
    Then \autoref{eq:kyp_lemma} is affine in $\bA$ and $\bB$, becoming
    \begin{align*}
        \sum_{j=1}^v\zeta(t)\begin{bmatrix}
            \he(\bP\bA^j)-\bQ & \bP\bB^j-\bS \\ 
            (\bP\bB^j)^T-\bS^T & -\bR
        \end{bmatrix}\preceq0.
    \end{align*}    
    Hence, \autoref{eq:polytopic_diss} implies \autoref{lem:kyp_lemma} holds for all admissible convex combinations of the polytopic system, establishing $QSR$-dissipativity of the operator $\scrG$. 
    $\hfill\blacksquare$
\end{pf}


The initial local controller gain for all agents is set to $[1000\;100]$.
To apply \autoref{alg:sparsity_promotion}, the parameters are chosen as $\bQ_{lq}{=}100\bI_{20}$, $\bR_{lq}{=}\bI_{20}$, $\epsilon_l{=}10^{-3}$, $\epsilon_p{=}\epsilon_d{=}10^{-4}$, $\rho{=}100$, and $\epsilon{=}10^{-3}$.
The weighting factor $\gamma$ is varied from 100 to 300 in increments of 5 for the weighted $\ell_1$ norm penalty, and from 335 to 1270 in increments of 5 for the cardinality penalty.
For the comparison, the approaches in \citep{lin2013design,lian2018sparsity} are implemented for the same system dynamics.
The same procedure as in \citep[Chap. 5]{lian2018sparsity} is used to compute the uncertainty input and output channels, with the $\cH_\infty$ constraint parameter $\gamma$ set to 1.25.
After obtaining the controllers, stability is evaluated over 200 randomly generated true dynamics of each agent, as well as over 256 vertex combinations constructed from the 16 vertex matrices associated with the stable and unstable agents. 
In the latter case, the stability analysis is performed under the simplifying assumption that all stable agents follow the same vertex matrix from the polytope corresponding to stable agents, and all unstable agents follow the same vertex matrix from the polyhedron corresponding to unstable agents.

\begin{figure}
\centering
    \resizebox{0.35\textwidth}{!}{
%
%
\definecolor{mycolor1}{rgb}{1.00000,0.00000,1.00000}%
\begin{tikzpicture}

\begin{axis}[%
width=4.521in,
height=3.566in,
at={(0.758in,0.481in)},
scale only axis,
xmin=10,
xmax=100,
xlabel style={font=\color{white!15!black}},
xlabel={Nonzero blocks},
ymin=3.6,
ymax=5.4,
ylabel style={font=\color{white!15!black}},
ylabel={log(LQ performance)},
axis background/.style={fill=white},
legend style={legend cell align=left, align=left, draw=white!15!black},
title style={font=\Huge},xlabel style={font={\LARGE}},ylabel style={font=\LARGE},legend style={font=\LARGE},
]
\addplot [color=blue, line width=2pt]
  table[row sep=crcr]{%
10	4.0550862344286\\
11	4.03965265365611\\
12	4.02635782659775\\
13	4.01763111738099\\
14	4.00255143834595\\
16	3.98926027593032\\
17	3.98699082288561\\
18	3.9823640976163\\
19	3.97094412425487\\
20	3.96981310694352\\
21	3.95728533979443\\
22	3.94835181129236\\
23	3.94429716493509\\
24	3.94393246066143\\
25	3.937326342892\\
26	3.93468538873306\\
27	3.93110280528427\\
28	3.92774111253655\\
31	3.91124462787491\\
32	3.90926275733222\\
33	3.90319155469665\\
36	3.89944174808986\\
38	3.88591118088716\\
40	3.87678060268225\\
41	3.87066062644793\\
43	3.86220137968744\\
45	3.85531169011503\\
46	3.85442462632589\\
49	3.84482868619649\\
50	3.84073802470689\\
51	3.83698001991257\\
52	3.83474336336283\\
53	3.83261732163984\\
54	3.82912282361238\\
55	3.82830873631287\\
56	3.82061787241424\\
59	3.81182204546474\\
60	3.81110118441474\\
61	3.81020450141196\\
62	3.80314647288685\\
63	3.80111999874421\\
64	3.79850810865046\\
66	3.79186786900265\\
68	3.7860407233079\\
69	3.78243168284984\\
71	3.77407695116997\\
72	3.77209356909993\\
74	3.76840181885719\\
75	3.76493739030185\\
76	3.76048546840474\\
77	3.75914212535849\\
78	3.75824567779186\\
79	3.75738233249396\\
81	3.75207966597307\\
82	3.7505369604329\\
83	3.74873956315725\\
84	3.74652600371686\\
86	3.74444894581149\\
89	3.74065862818743\\
90	3.73838696647881\\
92	3.73466685326843\\
93	3.7335632930228\\
95	3.73055963361447\\
96	3.72964050062652\\
98	3.72517949124215\\
99	3.72377423147931\\
100	3.72253002006926\\
};
\addlegendentry{Weighted $\ell_1$}

\addplot [color=red, line width=2pt]
  table[row sep=crcr]{%
10	4.0552127180848\\
12	4.02385288759839\\
14	4.01088923251645\\
15	4.00830814985249\\
16	4.00662899818818\\
18	3.97585267614314\\
19	3.97063738905257\\
20	3.96716596523398\\
21	3.9547483291047\\
22	3.9477434816643\\
23	3.93495628183033\\
24	3.93109445374382\\
26	3.93023652324354\\
27	3.92531584458817\\
28	3.92270377215867\\
30	3.9090243993222\\
31	3.90433220233411\\
32	3.89965422772132\\
33	3.89598127306593\\
34	3.88459100669051\\
35	3.87911295453131\\
36	3.87601382532046\\
37	3.87185573894488\\
38	3.86193279163506\\
39	3.86073564653248\\
40	3.85583194837011\\
41	3.85508937234151\\
42	3.85293126753197\\
44	3.84955368441937\\
45	3.84493157986207\\
47	3.84322018495616\\
49	3.83570465193669\\
50	3.83368414562415\\
51	3.83212917967835\\
54	3.82041205104047\\
56	3.81189157170422\\
57	3.80774961907091\\
60	3.8057556542424\\
61	3.80312322620918\\
63	3.79519815578107\\
65	3.79312209518231\\
66	3.79105715358565\\
68	3.78658079693182\\
69	3.78304147497173\\
71	3.78149947050392\\
74	3.77605295717116\\
75	3.77325045468548\\
77	3.77131916968331\\
79	3.76554186025662\\
81	3.76224234965431\\
82	3.7591647395605\\
83	3.75665473509618\\
85	3.75385906600723\\
86	3.75253161409678\\
89	3.74731231667809\\
90	3.74573131427236\\
93	3.74135070240692\\
94	3.73974700180369\\
95	3.73714495341252\\
98	3.73080483431315\\
100	3.72253243804856\\
};
\addlegendentry{Cardinality}

\addplot [color=green, line width=2pt]
  table[row sep=crcr]{%
10	3.95262718514412\\
12	3.9396969421175\\
13	3.92727371149625\\
14	3.92339717851427\\
15	3.91207209614713\\
16	3.90541658217309\\
17	3.89880683205928\\
18	3.86839242286348\\
19	3.83473785606105\\
20	3.84892864574935\\
21	3.8358422198049\\
22	3.8435409785902\\
23	3.83362218954064\\
24	3.77891168341173\\
25	3.72761045833448\\
26	3.71892000383491\\
27	3.71346754826628\\
28	3.70843497675717\\
29	3.70514833680682\\
30	3.71359167470885\\
31	3.7014259474803\\
32	3.70132135471127\\
34	3.69844322566527\\
35	3.75555681910643\\
36	3.6960720138213\\
37	3.78449477549113\\
38	3.80310273389057\\
39	3.70479577709618\\
50	3.72943876300034\\
51	3.72439929858609\\
52	3.72182719085434\\
53	3.72015664208839\\
55	3.68376824677688\\
60	3.68020420819424\\
61	3.679610638318\\
62	3.67923383647622\\
63	3.67886491311991\\
64	3.68085425801966\\
65	3.67814471752507\\
69	3.6758307260165\\
70	3.67550762414288\\
71	3.67448733647581\\
72	3.67375061398457\\
73	3.67374871922451\\
74	3.67316002388922\\
81	3.67443248366306\\
85	3.66753027214642\\
92	3.66312624597899\\
93	3.66242974890352\\
96	3.66073879670477\\
100	3.65856059229497\\
};
\addlegendentry{\citep{lin2013design}}

\addplot [color=mycolor1, line width=2pt]
  table[row sep=crcr]{%
10	5.30796093869264\\
12	5.23065356416098\\
14	5.16253345670726\\
16	5.11122318831065\\
18	5.05860434980754\\
20	5.0064447669238\\
22	5.20554659353388\\
24	5.10329251763006\\
26	5.19850137214776\\
28	5.00627007512247\\
30	4.70319818361317\\
32	5.11491467496988\\
34	4.95852800577665\\
36	4.66906769977314\\
38	4.67013920659182\\
40	4.84728071863054\\
42	4.66063406187455\\
44	4.64261978175273\\
46	4.6456793975917\\
48	4.63985752729577\\
50	4.71347466970449\\
52	4.62305480715065\\
54	4.6243949991051\\
56	4.62589981156162\\
58	4.61881288052034\\
60	4.62993710687095\\
62	4.60837746144255\\
64	4.60465716350576\\
66	4.60917215988308\\
68	4.6028325686938\\
70	4.58923056290569\\
72	4.58148945867465\\
74	4.57890400566926\\
76	4.57561920486112\\
78	4.56845501678894\\
80	4.56563647670106\\
82	4.56199211757337\\
84	4.55808140801168\\
86	4.56423876432286\\
88	4.55020482625152\\
90	4.54304569624837\\
92	4.53758051199868\\
94	4.52957288019461\\
96	4.52624220016681\\
98	4.52174027206677\\
};
\addlegendentry{\citep{lian2018sparsity}}

\end{axis}
\end{tikzpicture}%
}
    \vspace{-8pt}
    \caption{\gls{lq} performance index.} \label{fig:performance_analysis}
    \vspace*{-0.5\baselineskip} 
\end{figure}
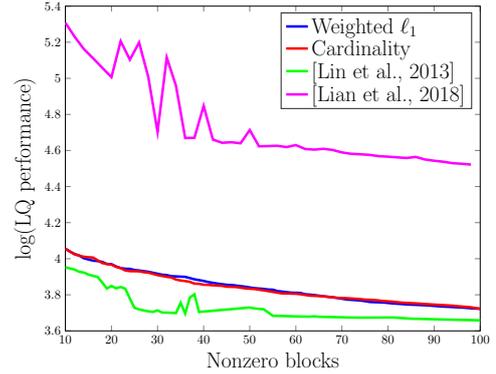


\begin{figure}
    \captionsetup[sub]{aboveskip=2pt, belowskip=0pt}
    \centering
    \subfloat[Weighted $\ell_1$]{\includegraphics[width=0.21\textwidth]{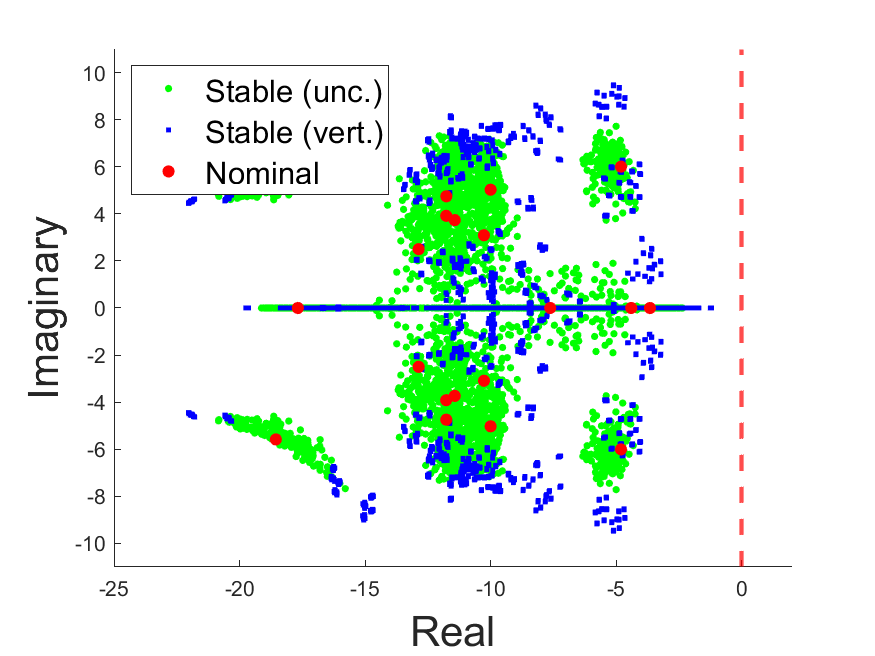}}
    \subfloat[Cardinality]{\includegraphics[width=0.21\textwidth]{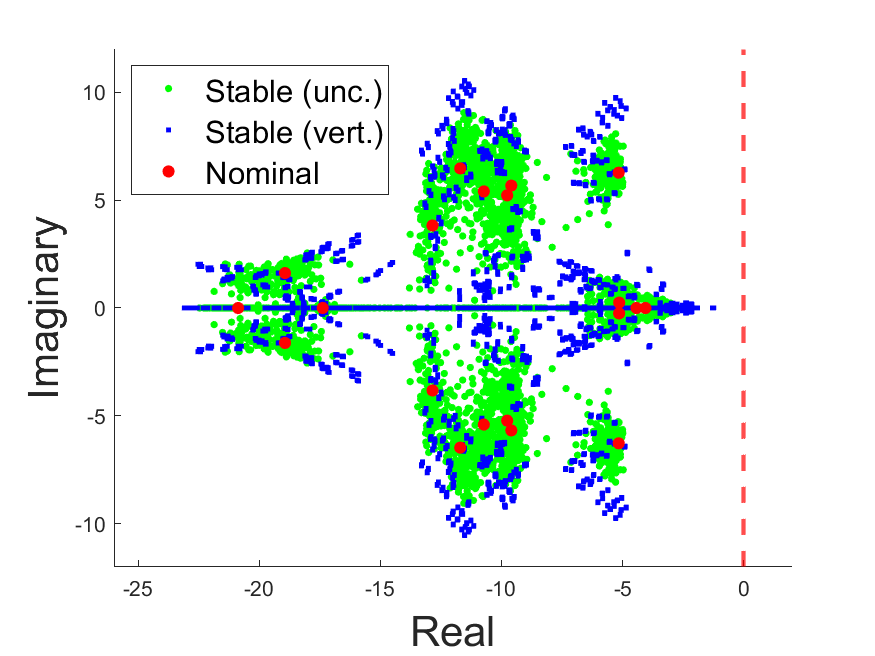}}
    \vspace{0.1pt}
    \subfloat[\citep{lin2013design}]{\includegraphics[width=0.21\textwidth]{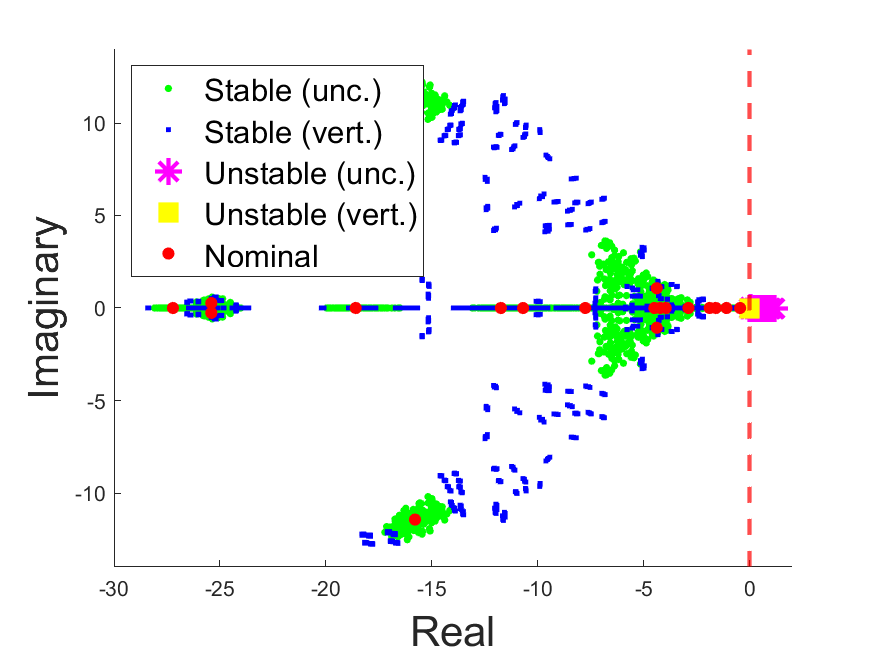}}
    \subfloat[\citep{lian2018sparsity}]{\includegraphics[width=0.21\textwidth]{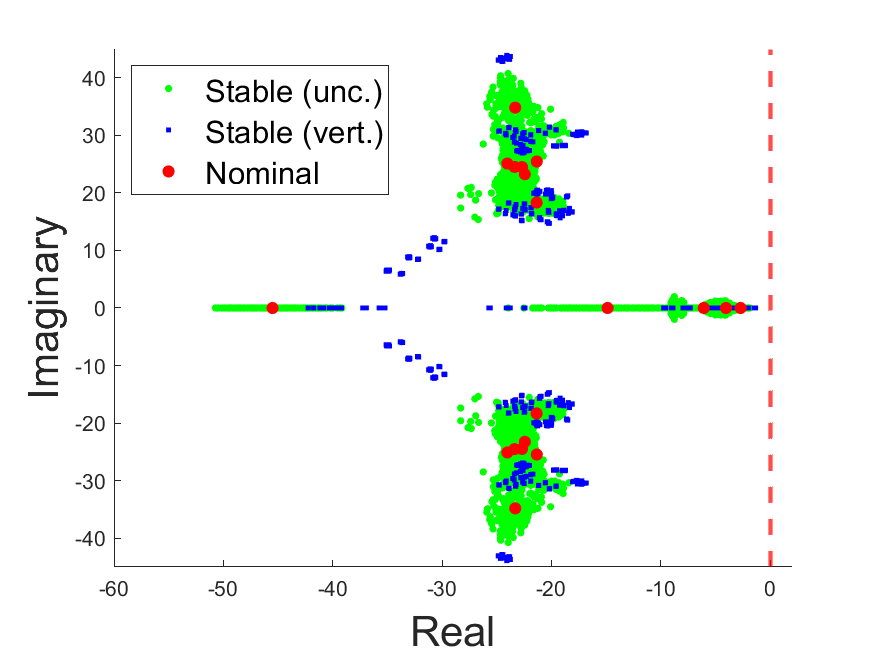}}
    \vspace{-8pt}
    \caption{Pole locations for ${\sum}_{i{,}j}\card(\norm{\bK_{ij}}_F){=}24$; ``(unc.)'' denotes poles from randomly generated uncertain dynamics of each agent, ``(vert.)'' denote poles from 256 vertices of polytopic uncertainty. The red dashed line indicates the boundary of the stable region.}
    \label{fig:pole_analysis}
    \vspace*{-0.5\baselineskip} 
\end{figure}

\begin{figure}
\centering
    \begin{subfigure}[t]{0.1\textwidth}
        \centering
        \resizebox{\textwidth}{!}{\begin{tikzpicture}
\definecolor{darkgreen}{rgb}{0,0.45,0}

\begin{axis}[%
width=4.521in,
height=3.566in,
at={(0.758in,0.481in)},
scale only axis,
xmin=-1.33946830265849,
xmax=11.3394683026585,
y dir=reverse,
ymin=-0.1,
ymax=10.1,
axis line style={draw=none},
ticks=none,
axis x line*=bottom,
axis y line*=left,
title style={font=\Huge},
]
\draw[black, line width=1pt]
  (axis cs:0,0) rectangle (axis cs:10,10);
\draw[fill=darkgreen, draw=none] (axis cs:0,0) rectangle (axis cs:1,1);
\draw[fill=darkgreen, draw=none] (axis cs:1,0) rectangle (axis cs:2,1);
\draw[fill=darkgreen, draw=none] (axis cs:2,0) rectangle (axis cs:3,1);
\draw[fill=white, draw=none] (axis cs:3,0) rectangle (axis cs:4,1);
\draw[fill=darkgreen, draw=none] (axis cs:4,0) rectangle (axis cs:5,1);
\draw[fill=white, draw=none] (axis cs:5,0) rectangle (axis cs:6,1);
\draw[fill=white, draw=none] (axis cs:6,0) rectangle (axis cs:7,1);
\draw[fill=white, draw=none] (axis cs:7,0) rectangle (axis cs:8,1);
\draw[fill=darkgreen, draw=none] (axis cs:8,0) rectangle (axis cs:9,1);
\draw[fill=white, draw=none] (axis cs:9,0) rectangle (axis cs:10,1);
\draw[fill=darkgreen, draw=none] (axis cs:0,1) rectangle (axis cs:1,2);
\draw[fill=darkgreen, draw=none] (axis cs:1,1) rectangle (axis cs:2,2);
\draw[fill=darkgreen, draw=none] (axis cs:2,1) rectangle (axis cs:3,2);
\draw[fill=darkgreen, draw=none] (axis cs:3,1) rectangle (axis cs:4,2);
\draw[fill=darkgreen, draw=none] (axis cs:4,1) rectangle (axis cs:5,2);
\draw[fill=darkgreen, draw=none] (axis cs:5,1) rectangle (axis cs:6,2);
\draw[fill=white, draw=none] (axis cs:6,1) rectangle (axis cs:7,2);
\draw[fill=darkgreen, draw=none] (axis cs:7,1) rectangle (axis cs:8,2);
\draw[fill=white, draw=none] (axis cs:8,1) rectangle (axis cs:9,2);
\draw[fill=white, draw=none] (axis cs:9,1) rectangle (axis cs:10,2);
\draw[fill=darkgreen, draw=none] (axis cs:0,2) rectangle (axis cs:1,3);
\draw[fill=darkgreen, draw=none] (axis cs:1,2) rectangle (axis cs:2,3);
\draw[fill=darkgreen, draw=none] (axis cs:2,2) rectangle (axis cs:3,3);
\draw[fill=white, draw=none] (axis cs:3,2) rectangle (axis cs:4,3);
\draw[fill=darkgreen, draw=none] (axis cs:4,2) rectangle (axis cs:5,3);
\draw[fill=white, draw=none] (axis cs:5,2) rectangle (axis cs:6,3);
\draw[fill=white, draw=none] (axis cs:6,2) rectangle (axis cs:7,3);
\draw[fill=white, draw=none] (axis cs:7,2) rectangle (axis cs:8,3);
\draw[fill=darkgreen, draw=none] (axis cs:8,2) rectangle (axis cs:9,3);
\draw[fill=white, draw=none] (axis cs:9,2) rectangle (axis cs:10,3);
\draw[fill=darkgreen, draw=none] (axis cs:0,3) rectangle (axis cs:1,4);
\draw[fill=darkgreen, draw=none] (axis cs:1,3) rectangle (axis cs:2,4);
\draw[fill=darkgreen, draw=none] (axis cs:2,3) rectangle (axis cs:3,4);
\draw[fill=darkgreen, draw=none] (axis cs:3,3) rectangle (axis cs:4,4);
\draw[fill=darkgreen, draw=none] (axis cs:4,3) rectangle (axis cs:5,4);
\draw[fill=white, draw=none] (axis cs:5,3) rectangle (axis cs:6,4);
\draw[fill=white, draw=none] (axis cs:6,3) rectangle (axis cs:7,4);
\draw[fill=white, draw=none] (axis cs:7,3) rectangle (axis cs:8,4);
\draw[fill=white, draw=none] (axis cs:8,3) rectangle (axis cs:9,4);
\draw[fill=white, draw=none] (axis cs:9,3) rectangle (axis cs:10,4);
\draw[fill=darkgreen, draw=none] (axis cs:0,4) rectangle (axis cs:1,5);
\draw[fill=darkgreen, draw=none] (axis cs:1,4) rectangle (axis cs:2,5);
\draw[fill=darkgreen, draw=none] (axis cs:2,4) rectangle (axis cs:3,5);
\draw[fill=darkgreen, draw=none] (axis cs:3,4) rectangle (axis cs:4,5);
\draw[fill=darkgreen, draw=none] (axis cs:4,4) rectangle (axis cs:5,5);
\draw[fill=white, draw=none] (axis cs:5,4) rectangle (axis cs:6,5);
\draw[fill=white, draw=none] (axis cs:6,4) rectangle (axis cs:7,5);
\draw[fill=darkgreen, draw=none] (axis cs:7,4) rectangle (axis cs:8,5);
\draw[fill=white, draw=none] (axis cs:8,4) rectangle (axis cs:9,5);
\draw[fill=white, draw=none] (axis cs:9,4) rectangle (axis cs:10,5);
\draw[fill=white, draw=none] (axis cs:0,5) rectangle (axis cs:1,6);
\draw[fill=darkgreen, draw=none] (axis cs:1,5) rectangle (axis cs:2,6);
\draw[fill=white, draw=none] (axis cs:2,5) rectangle (axis cs:3,6);
\draw[fill=darkgreen, draw=none] (axis cs:3,5) rectangle (axis cs:4,6);
\draw[fill=darkgreen, draw=none] (axis cs:4,5) rectangle (axis cs:5,6);
\draw[fill=darkgreen, draw=none] (axis cs:5,5) rectangle (axis cs:6,6);
\draw[fill=white, draw=none] (axis cs:6,5) rectangle (axis cs:7,6);
\draw[fill=white, draw=none] (axis cs:7,5) rectangle (axis cs:8,6);
\draw[fill=white, draw=none] (axis cs:8,5) rectangle (axis cs:9,6);
\draw[fill=white, draw=none] (axis cs:9,5) rectangle (axis cs:10,6);
\draw[fill=darkgreen, draw=none] (axis cs:0,6) rectangle (axis cs:1,7);
\draw[fill=darkgreen, draw=none] (axis cs:1,6) rectangle (axis cs:2,7);
\draw[fill=darkgreen, draw=none] (axis cs:2,6) rectangle (axis cs:3,7);
\draw[fill=white, draw=none] (axis cs:3,6) rectangle (axis cs:4,7);
\draw[fill=darkgreen, draw=none] (axis cs:4,6) rectangle (axis cs:5,7);
\draw[fill=white, draw=none] (axis cs:5,6) rectangle (axis cs:6,7);
\draw[fill=darkgreen, draw=none] (axis cs:6,6) rectangle (axis cs:7,7);
\draw[fill=white, draw=none] (axis cs:7,6) rectangle (axis cs:8,7);
\draw[fill=white, draw=none] (axis cs:8,6) rectangle (axis cs:9,7);
\draw[fill=white, draw=none] (axis cs:9,6) rectangle (axis cs:10,7);
\draw[fill=darkgreen, draw=none] (axis cs:0,7) rectangle (axis cs:1,8);
\draw[fill=darkgreen, draw=none] (axis cs:1,7) rectangle (axis cs:2,8);
\draw[fill=darkgreen, draw=none] (axis cs:2,7) rectangle (axis cs:3,8);
\draw[fill=white, draw=none] (axis cs:3,7) rectangle (axis cs:4,8);
\draw[fill=darkgreen, draw=none] (axis cs:4,7) rectangle (axis cs:5,8);
\draw[fill=white, draw=none] (axis cs:5,7) rectangle (axis cs:6,8);
\draw[fill=white, draw=none] (axis cs:6,7) rectangle (axis cs:7,8);
\draw[fill=darkgreen, draw=none] (axis cs:7,7) rectangle (axis cs:8,8);
\draw[fill=white, draw=none] (axis cs:8,7) rectangle (axis cs:9,8);
\draw[fill=white, draw=none] (axis cs:9,7) rectangle (axis cs:10,8);
\draw[fill=darkgreen, draw=none] (axis cs:0,8) rectangle (axis cs:1,9);
\draw[fill=white, draw=none] (axis cs:1,8) rectangle (axis cs:2,9);
\draw[fill=darkgreen, draw=none] (axis cs:2,8) rectangle (axis cs:3,9);
\draw[fill=white, draw=none] (axis cs:3,8) rectangle (axis cs:4,9);
\draw[fill=darkgreen, draw=none] (axis cs:4,8) rectangle (axis cs:5,9);
\draw[fill=white, draw=none] (axis cs:5,8) rectangle (axis cs:6,9);
\draw[fill=white, draw=none] (axis cs:6,8) rectangle (axis cs:7,9);
\draw[fill=white, draw=none] (axis cs:7,8) rectangle (axis cs:8,9);
\draw[fill=darkgreen, draw=none] (axis cs:8,8) rectangle (axis cs:9,9);
\draw[fill=white, draw=none] (axis cs:9,8) rectangle (axis cs:10,9);
\draw[fill=darkgreen, draw=none] (axis cs:0,9) rectangle (axis cs:1,10);
\draw[fill=white, draw=none] (axis cs:1,9) rectangle (axis cs:2,10);
\draw[fill=darkgreen, draw=none] (axis cs:2,9) rectangle (axis cs:3,10);
\draw[fill=white, draw=none] (axis cs:3,9) rectangle (axis cs:4,10);
\draw[fill=darkgreen, draw=none] (axis cs:4,9) rectangle (axis cs:5,10);
\draw[fill=white, draw=none] (axis cs:5,9) rectangle (axis cs:6,10);
\draw[fill=white, draw=none] (axis cs:6,9) rectangle (axis cs:7,10);
\draw[fill=white, draw=none] (axis cs:7,9) rectangle (axis cs:8,10);
\draw[fill=white, draw=none] (axis cs:8,9) rectangle (axis cs:9,10);
\draw[fill=darkgreen, draw=none] (axis cs:9,9) rectangle (axis cs:10,10);
\end{axis}
\end{tikzpicture}%
}
    \end{subfigure}
    \begin{subfigure}[t]{0.1\textwidth}
        \centering
        \resizebox{\textwidth}{!}{\begin{tikzpicture}
\definecolor{darkgreen}{rgb}{0,0.45,0}

\begin{axis}[%
width=4.521in,
height=3.566in,
at={(0.758in,0.481in)},
scale only axis,
xmin=-1.33946830265849,
xmax=11.3394683026585,
y dir=reverse,
ymin=-0.1,
ymax=10.1,
axis line style={draw=none},
ticks=none,
axis x line*=bottom,
axis y line*=left,
title style={font=\Huge},
]
\draw[black, line width=1pt]
  (axis cs:0,0) rectangle (axis cs:10,10);
\draw[fill=darkgreen, draw=none] (axis cs:0,0) rectangle (axis cs:1,1);
\draw[fill=darkgreen, draw=none] (axis cs:1,0) rectangle (axis cs:2,1);
\draw[fill=darkgreen, draw=none] (axis cs:2,0) rectangle (axis cs:3,1);
\draw[fill=white, draw=none] (axis cs:3,0) rectangle (axis cs:4,1);
\draw[fill=darkgreen, draw=none] (axis cs:4,0) rectangle (axis cs:5,1);
\draw[fill=white, draw=none] (axis cs:5,0) rectangle (axis cs:6,1);
\draw[fill=darkgreen, draw=none] (axis cs:6,0) rectangle (axis cs:7,1);
\draw[fill=darkgreen, draw=none] (axis cs:7,0) rectangle (axis cs:8,1);
\draw[fill=darkgreen, draw=none] (axis cs:8,0) rectangle (axis cs:9,1);
\draw[fill=white, draw=none] (axis cs:9,0) rectangle (axis cs:10,1);
\draw[fill=darkgreen, draw=none] (axis cs:0,1) rectangle (axis cs:1,2);
\draw[fill=darkgreen, draw=none] (axis cs:1,1) rectangle (axis cs:2,2);
\draw[fill=darkgreen, draw=none] (axis cs:2,1) rectangle (axis cs:3,2);
\draw[fill=darkgreen, draw=none] (axis cs:3,1) rectangle (axis cs:4,2);
\draw[fill=darkgreen, draw=none] (axis cs:4,1) rectangle (axis cs:5,2);
\draw[fill=darkgreen, draw=none] (axis cs:5,1) rectangle (axis cs:6,2);
\draw[fill=white, draw=none] (axis cs:6,1) rectangle (axis cs:7,2);
\draw[fill=darkgreen, draw=none] (axis cs:7,1) rectangle (axis cs:8,2);
\draw[fill=white, draw=none] (axis cs:8,1) rectangle (axis cs:9,2);
\draw[fill=white, draw=none] (axis cs:9,1) rectangle (axis cs:10,2);
\draw[fill=darkgreen, draw=none] (axis cs:0,2) rectangle (axis cs:1,3);
\draw[fill=darkgreen, draw=none] (axis cs:1,2) rectangle (axis cs:2,3);
\draw[fill=darkgreen, draw=none] (axis cs:2,2) rectangle (axis cs:3,3);
\draw[fill=white, draw=none] (axis cs:3,2) rectangle (axis cs:4,3);
\draw[fill=darkgreen, draw=none] (axis cs:4,2) rectangle (axis cs:5,3);
\draw[fill=white, draw=none] (axis cs:5,2) rectangle (axis cs:6,3);
\draw[fill=darkgreen, draw=none] (axis cs:6,2) rectangle (axis cs:7,3);
\draw[fill=darkgreen, draw=none] (axis cs:7,2) rectangle (axis cs:8,3);
\draw[fill=darkgreen, draw=none] (axis cs:8,2) rectangle (axis cs:9,3);
\draw[fill=white, draw=none] (axis cs:9,2) rectangle (axis cs:10,3);
\draw[fill=white, draw=none] (axis cs:0,3) rectangle (axis cs:1,4);
\draw[fill=darkgreen, draw=none] (axis cs:1,3) rectangle (axis cs:2,4);
\draw[fill=white, draw=none] (axis cs:2,3) rectangle (axis cs:3,4);
\draw[fill=darkgreen, draw=none] (axis cs:3,3) rectangle (axis cs:4,4);
\draw[fill=darkgreen, draw=none] (axis cs:4,3) rectangle (axis cs:5,4);
\draw[fill=darkgreen, draw=none] (axis cs:5,3) rectangle (axis cs:6,4);
\draw[fill=white, draw=none] (axis cs:6,3) rectangle (axis cs:7,4);
\draw[fill=white, draw=none] (axis cs:7,3) rectangle (axis cs:8,4);
\draw[fill=white, draw=none] (axis cs:8,3) rectangle (axis cs:9,4);
\draw[fill=white, draw=none] (axis cs:9,3) rectangle (axis cs:10,4);
\draw[fill=darkgreen, draw=none] (axis cs:0,4) rectangle (axis cs:1,5);
\draw[fill=darkgreen, draw=none] (axis cs:1,4) rectangle (axis cs:2,5);
\draw[fill=darkgreen, draw=none] (axis cs:2,4) rectangle (axis cs:3,5);
\draw[fill=darkgreen, draw=none] (axis cs:3,4) rectangle (axis cs:4,5);
\draw[fill=darkgreen, draw=none] (axis cs:4,4) rectangle (axis cs:5,5);
\draw[fill=darkgreen, draw=none] (axis cs:5,4) rectangle (axis cs:6,5);
\draw[fill=darkgreen, draw=none] (axis cs:6,4) rectangle (axis cs:7,5);
\draw[fill=darkgreen, draw=none] (axis cs:7,4) rectangle (axis cs:8,5);
\draw[fill=darkgreen, draw=none] (axis cs:8,4) rectangle (axis cs:9,5);
\draw[fill=white, draw=none] (axis cs:9,4) rectangle (axis cs:10,5);
\draw[fill=white, draw=none] (axis cs:0,5) rectangle (axis cs:1,6);
\draw[fill=darkgreen, draw=none] (axis cs:1,5) rectangle (axis cs:2,6);
\draw[fill=white, draw=none] (axis cs:2,5) rectangle (axis cs:3,6);
\draw[fill=darkgreen, draw=none] (axis cs:3,5) rectangle (axis cs:4,6);
\draw[fill=white, draw=none] (axis cs:4,5) rectangle (axis cs:5,6);
\draw[fill=darkgreen, draw=none] (axis cs:5,5) rectangle (axis cs:6,6);
\draw[fill=white, draw=none] (axis cs:6,5) rectangle (axis cs:7,6);
\draw[fill=white, draw=none] (axis cs:7,5) rectangle (axis cs:8,6);
\draw[fill=white, draw=none] (axis cs:8,5) rectangle (axis cs:9,6);
\draw[fill=white, draw=none] (axis cs:9,5) rectangle (axis cs:10,6);
\draw[fill=white, draw=none] (axis cs:0,6) rectangle (axis cs:1,7);
\draw[fill=white, draw=none] (axis cs:1,6) rectangle (axis cs:2,7);
\draw[fill=darkgreen, draw=none] (axis cs:2,6) rectangle (axis cs:3,7);
\draw[fill=white, draw=none] (axis cs:3,6) rectangle (axis cs:4,7);
\draw[fill=darkgreen, draw=none] (axis cs:4,6) rectangle (axis cs:5,7);
\draw[fill=white, draw=none] (axis cs:5,6) rectangle (axis cs:6,7);
\draw[fill=darkgreen, draw=none] (axis cs:6,6) rectangle (axis cs:7,7);
\draw[fill=white, draw=none] (axis cs:7,6) rectangle (axis cs:8,7);
\draw[fill=white, draw=none] (axis cs:8,6) rectangle (axis cs:9,7);
\draw[fill=white, draw=none] (axis cs:9,6) rectangle (axis cs:10,7);
\draw[fill=darkgreen, draw=none] (axis cs:0,7) rectangle (axis cs:1,8);
\draw[fill=darkgreen, draw=none] (axis cs:1,7) rectangle (axis cs:2,8);
\draw[fill=darkgreen, draw=none] (axis cs:2,7) rectangle (axis cs:3,8);
\draw[fill=white, draw=none] (axis cs:3,7) rectangle (axis cs:4,8);
\draw[fill=darkgreen, draw=none] (axis cs:4,7) rectangle (axis cs:5,8);
\draw[fill=white, draw=none] (axis cs:5,7) rectangle (axis cs:6,8);
\draw[fill=white, draw=none] (axis cs:6,7) rectangle (axis cs:7,8);
\draw[fill=darkgreen, draw=none] (axis cs:7,7) rectangle (axis cs:8,8);
\draw[fill=white, draw=none] (axis cs:8,7) rectangle (axis cs:9,8);
\draw[fill=white, draw=none] (axis cs:9,7) rectangle (axis cs:10,8);
\draw[fill=darkgreen, draw=none] (axis cs:0,8) rectangle (axis cs:1,9);
\draw[fill=white, draw=none] (axis cs:1,8) rectangle (axis cs:2,9);
\draw[fill=darkgreen, draw=none] (axis cs:2,8) rectangle (axis cs:3,9);
\draw[fill=white, draw=none] (axis cs:3,8) rectangle (axis cs:4,9);
\draw[fill=darkgreen, draw=none] (axis cs:4,8) rectangle (axis cs:5,9);
\draw[fill=white, draw=none] (axis cs:5,8) rectangle (axis cs:6,9);
\draw[fill=white, draw=none] (axis cs:6,8) rectangle (axis cs:7,9);
\draw[fill=white, draw=none] (axis cs:7,8) rectangle (axis cs:8,9);
\draw[fill=darkgreen, draw=none] (axis cs:8,8) rectangle (axis cs:9,9);
\draw[fill=white, draw=none] (axis cs:9,8) rectangle (axis cs:10,9);
\draw[fill=white, draw=none] (axis cs:0,9) rectangle (axis cs:1,10);
\draw[fill=white, draw=none] (axis cs:1,9) rectangle (axis cs:2,10);
\draw[fill=white, draw=none] (axis cs:2,9) rectangle (axis cs:3,10);
\draw[fill=white, draw=none] (axis cs:3,9) rectangle (axis cs:4,10);
\draw[fill=white, draw=none] (axis cs:4,9) rectangle (axis cs:5,10);
\draw[fill=white, draw=none] (axis cs:5,9) rectangle (axis cs:6,10);
\draw[fill=white, draw=none] (axis cs:6,9) rectangle (axis cs:7,10);
\draw[fill=white, draw=none] (axis cs:7,9) rectangle (axis cs:8,10);
\draw[fill=white, draw=none] (axis cs:8,9) rectangle (axis cs:9,10);
\draw[fill=darkgreen, draw=none] (axis cs:9,9) rectangle (axis cs:10,10);
\end{axis}
\end{tikzpicture}%
}
    \end{subfigure}
    \begin{subfigure}[t]{0.1\textwidth}
        \centering
        \resizebox{\textwidth}{!}{\begin{tikzpicture}
\definecolor{darkgreen}{rgb}{0,0.45,0}

\begin{axis}[%
width=4.521in,
height=3.566in,
at={(0.758in,0.481in)},
scale only axis,
xmin=-1.33946830265849,
xmax=11.3394683026585,
y dir=reverse,
ymin=-.1,
ymax=10.1,
axis line style={draw=none},
ticks=none,
axis x line*=bottom,
axis y line*=left,
title style={font=\Huge},
]
\draw[black, line width=1pt]
  (axis cs:0,0) rectangle (axis cs:10,10);
\draw[fill=darkgreen, draw=none] (axis cs:0,0) rectangle (axis cs:1,1);
\draw[fill=darkgreen, draw=none] (axis cs:1,0) rectangle (axis cs:2,1);
\draw[fill=darkgreen, draw=none] (axis cs:2,0) rectangle (axis cs:3,1);
\draw[fill=darkgreen, draw=none] (axis cs:3,0) rectangle (axis cs:4,1);
\draw[fill=darkgreen, draw=none] (axis cs:4,0) rectangle (axis cs:5,1);
\draw[fill=white, draw=none] (axis cs:5,0) rectangle (axis cs:6,1);
\draw[fill=darkgreen, draw=none] (axis cs:6,0) rectangle (axis cs:7,1);
\draw[fill=darkgreen, draw=none] (axis cs:7,0) rectangle (axis cs:8,1);
\draw[fill=darkgreen, draw=none] (axis cs:8,0) rectangle (axis cs:9,1);
\draw[fill=darkgreen, draw=none] (axis cs:9,0) rectangle (axis cs:10,1);
\draw[fill=darkgreen, draw=none] (axis cs:0,1) rectangle (axis cs:1,2);
\draw[fill=darkgreen, draw=none] (axis cs:1,1) rectangle (axis cs:2,2);
\draw[fill=darkgreen, draw=none] (axis cs:2,1) rectangle (axis cs:3,2);
\draw[fill=darkgreen, draw=none] (axis cs:3,1) rectangle (axis cs:4,2);
\draw[fill=darkgreen, draw=none] (axis cs:4,1) rectangle (axis cs:5,2);
\draw[fill=darkgreen, draw=none] (axis cs:5,1) rectangle (axis cs:6,2);
\draw[fill=darkgreen, draw=none] (axis cs:6,1) rectangle (axis cs:7,2);
\draw[fill=darkgreen, draw=none] (axis cs:7,1) rectangle (axis cs:8,2);
\draw[fill=darkgreen, draw=none] (axis cs:8,1) rectangle (axis cs:9,2);
\draw[fill=darkgreen, draw=none] (axis cs:9,1) rectangle (axis cs:10,2);
\draw[fill=darkgreen, draw=none] (axis cs:0,2) rectangle (axis cs:1,3);
\draw[fill=darkgreen, draw=none] (axis cs:1,2) rectangle (axis cs:2,3);
\draw[fill=darkgreen, draw=none] (axis cs:2,2) rectangle (axis cs:3,3);
\draw[fill=darkgreen, draw=none] (axis cs:3,2) rectangle (axis cs:4,3);
\draw[fill=darkgreen, draw=none] (axis cs:4,2) rectangle (axis cs:5,3);
\draw[fill=white, draw=none] (axis cs:5,2) rectangle (axis cs:6,3);
\draw[fill=darkgreen, draw=none] (axis cs:6,2) rectangle (axis cs:7,3);
\draw[fill=darkgreen, draw=none] (axis cs:7,2) rectangle (axis cs:8,3);
\draw[fill=darkgreen, draw=none] (axis cs:8,2) rectangle (axis cs:9,3);
\draw[fill=darkgreen, draw=none] (axis cs:9,2) rectangle (axis cs:10,3);
\draw[fill=white, draw=none] (axis cs:0,3) rectangle (axis cs:1,4);
\draw[fill=darkgreen, draw=none] (axis cs:1,3) rectangle (axis cs:2,4);
\draw[fill=white, draw=none] (axis cs:2,3) rectangle (axis cs:3,4);
\draw[fill=darkgreen, draw=none] (axis cs:3,3) rectangle (axis cs:4,4);
\draw[fill=darkgreen, draw=none] (axis cs:4,3) rectangle (axis cs:5,4);
\draw[fill=darkgreen, draw=none] (axis cs:5,3) rectangle (axis cs:6,4);
\draw[fill=white, draw=none] (axis cs:6,3) rectangle (axis cs:7,4);
\draw[fill=darkgreen, draw=none] (axis cs:7,3) rectangle (axis cs:8,4);
\draw[fill=white, draw=none] (axis cs:8,3) rectangle (axis cs:9,4);
\draw[fill=white, draw=none] (axis cs:9,3) rectangle (axis cs:10,4);
\draw[fill=darkgreen, draw=none] (axis cs:0,4) rectangle (axis cs:1,5);
\draw[fill=darkgreen, draw=none] (axis cs:1,4) rectangle (axis cs:2,5);
\draw[fill=darkgreen, draw=none] (axis cs:2,4) rectangle (axis cs:3,5);
\draw[fill=darkgreen, draw=none] (axis cs:3,4) rectangle (axis cs:4,5);
\draw[fill=darkgreen, draw=none] (axis cs:4,4) rectangle (axis cs:5,5);
\draw[fill=darkgreen, draw=none] (axis cs:5,4) rectangle (axis cs:6,5);
\draw[fill=darkgreen, draw=none] (axis cs:6,4) rectangle (axis cs:7,5);
\draw[fill=darkgreen, draw=none] (axis cs:7,4) rectangle (axis cs:8,5);
\draw[fill=darkgreen, draw=none] (axis cs:8,4) rectangle (axis cs:9,5);
\draw[fill=darkgreen, draw=none] (axis cs:9,4) rectangle (axis cs:10,5);
\draw[fill=white, draw=none] (axis cs:0,5) rectangle (axis cs:1,6);
\draw[fill=white, draw=none] (axis cs:1,5) rectangle (axis cs:2,6);
\draw[fill=white, draw=none] (axis cs:2,5) rectangle (axis cs:3,6);
\draw[fill=white, draw=none] (axis cs:3,5) rectangle (axis cs:4,6);
\draw[fill=white, draw=none] (axis cs:4,5) rectangle (axis cs:5,6);
\draw[fill=darkgreen, draw=none] (axis cs:5,5) rectangle (axis cs:6,6);
\draw[fill=white, draw=none] (axis cs:6,5) rectangle (axis cs:7,6);
\draw[fill=white, draw=none] (axis cs:7,5) rectangle (axis cs:8,6);
\draw[fill=white, draw=none] (axis cs:8,5) rectangle (axis cs:9,6);
\draw[fill=white, draw=none] (axis cs:9,5) rectangle (axis cs:10,6);
\draw[fill=white, draw=none] (axis cs:0,6) rectangle (axis cs:1,7);
\draw[fill=white, draw=none] (axis cs:1,6) rectangle (axis cs:2,7);
\draw[fill=white, draw=none] (axis cs:2,6) rectangle (axis cs:3,7);
\draw[fill=white, draw=none] (axis cs:3,6) rectangle (axis cs:4,7);
\draw[fill=darkgreen, draw=none] (axis cs:4,6) rectangle (axis cs:5,7);
\draw[fill=white, draw=none] (axis cs:5,6) rectangle (axis cs:6,7);
\draw[fill=darkgreen, draw=none] (axis cs:6,6) rectangle (axis cs:7,7);
\draw[fill=white, draw=none] (axis cs:7,6) rectangle (axis cs:8,7);
\draw[fill=white, draw=none] (axis cs:8,6) rectangle (axis cs:9,7);
\draw[fill=white, draw=none] (axis cs:9,6) rectangle (axis cs:10,7);
\draw[fill=white, draw=none] (axis cs:0,7) rectangle (axis cs:1,8);
\draw[fill=white, draw=none] (axis cs:1,7) rectangle (axis cs:2,8);
\draw[fill=white, draw=none] (axis cs:2,7) rectangle (axis cs:3,8);
\draw[fill=white, draw=none] (axis cs:3,7) rectangle (axis cs:4,8);
\draw[fill=darkgreen, draw=none] (axis cs:4,7) rectangle (axis cs:5,8);
\draw[fill=white, draw=none] (axis cs:5,7) rectangle (axis cs:6,8);
\draw[fill=white, draw=none] (axis cs:6,7) rectangle (axis cs:7,8);
\draw[fill=darkgreen, draw=none] (axis cs:7,7) rectangle (axis cs:8,8);
\draw[fill=white, draw=none] (axis cs:8,7) rectangle (axis cs:9,8);
\draw[fill=white, draw=none] (axis cs:9,7) rectangle (axis cs:10,8);
\draw[fill=white, draw=none] (axis cs:0,8) rectangle (axis cs:1,9);
\draw[fill=white, draw=none] (axis cs:1,8) rectangle (axis cs:2,9);
\draw[fill=white, draw=none] (axis cs:2,8) rectangle (axis cs:3,9);
\draw[fill=white, draw=none] (axis cs:3,8) rectangle (axis cs:4,9);
\draw[fill=white, draw=none] (axis cs:4,8) rectangle (axis cs:5,9);
\draw[fill=white, draw=none] (axis cs:5,8) rectangle (axis cs:6,9);
\draw[fill=white, draw=none] (axis cs:6,8) rectangle (axis cs:7,9);
\draw[fill=white, draw=none] (axis cs:7,8) rectangle (axis cs:8,9);
\draw[fill=darkgreen, draw=none] (axis cs:8,8) rectangle (axis cs:9,9);
\draw[fill=white, draw=none] (axis cs:9,8) rectangle (axis cs:10,9);
\draw[fill=white, draw=none] (axis cs:0,9) rectangle (axis cs:1,10);
\draw[fill=white, draw=none] (axis cs:1,9) rectangle (axis cs:2,10);
\draw[fill=white, draw=none] (axis cs:2,9) rectangle (axis cs:3,10);
\draw[fill=white, draw=none] (axis cs:3,9) rectangle (axis cs:4,10);
\draw[fill=white, draw=none] (axis cs:4,9) rectangle (axis cs:5,10);
\draw[fill=white, draw=none] (axis cs:5,9) rectangle (axis cs:6,10);
\draw[fill=white, draw=none] (axis cs:6,9) rectangle (axis cs:7,10);
\draw[fill=white, draw=none] (axis cs:7,9) rectangle (axis cs:8,10);
\draw[fill=white, draw=none] (axis cs:8,9) rectangle (axis cs:9,10);
\draw[fill=darkgreen, draw=none] (axis cs:9,9) rectangle (axis cs:10,10);
\end{axis}
\end{tikzpicture}%
}
    \end{subfigure}
    \begin{subfigure}[t]{0.1\textwidth}
        \centering
        \resizebox{\textwidth}{!}{
%
%
\begin{tikzpicture}
\definecolor{darkgreen}{rgb}{0,0.45,0}

\begin{axis}[%
width=4.521in,
height=3.566in,
at={(0.758in,0.481in)},
scale only axis,
xmin=-1.33946830265849,
xmax=11.3394683026585,
y dir=reverse,
ymin=-.1,
ymax=10.1,
axis line style={draw=none},
ticks=none,
axis x line*=bottom,
axis y line*=left,
title style={font=\Huge},
]
\draw[black, line width=1pt]
  (axis cs:0,0) rectangle (axis cs:10,10);
\draw[fill=darkgreen, draw=none] (axis cs:0,0) rectangle (axis cs:1,1);
\draw[fill=white, draw=none] (axis cs:1,0) rectangle (axis cs:2,1);
\draw[fill=white, draw=none] (axis cs:2,0) rectangle (axis cs:3,1);
\draw[fill=darkgreen, draw=none] (axis cs:3,0) rectangle (axis cs:4,1);
\draw[fill=darkgreen, draw=none] (axis cs:4,0) rectangle (axis cs:5,1);
\draw[fill=white, draw=none] (axis cs:5,0) rectangle (axis cs:6,1);
\draw[fill=darkgreen, draw=none] (axis cs:6,0) rectangle (axis cs:7,1);
\draw[fill=darkgreen, draw=none] (axis cs:7,0) rectangle (axis cs:8,1);
\draw[fill=darkgreen, draw=none] (axis cs:8,0) rectangle (axis cs:9,1);
\draw[fill=white, draw=none] (axis cs:9,0) rectangle (axis cs:10,1);
\draw[fill=white, draw=none] (axis cs:0,1) rectangle (axis cs:1,2);
\draw[fill=darkgreen, draw=none] (axis cs:1,1) rectangle (axis cs:2,2);
\draw[fill=white, draw=none] (axis cs:2,1) rectangle (axis cs:3,2);
\draw[fill=darkgreen, draw=none] (axis cs:3,1) rectangle (axis cs:4,2);
\draw[fill=darkgreen, draw=none] (axis cs:4,1) rectangle (axis cs:5,2);
\draw[fill=white, draw=none] (axis cs:5,1) rectangle (axis cs:6,2);
\draw[fill=darkgreen, draw=none] (axis cs:6,1) rectangle (axis cs:7,2);
\draw[fill=darkgreen, draw=none] (axis cs:7,1) rectangle (axis cs:8,2);
\draw[fill=white, draw=none] (axis cs:8,1) rectangle (axis cs:9,2);
\draw[fill=white, draw=none] (axis cs:9,1) rectangle (axis cs:10,2);
\draw[fill=white, draw=none] (axis cs:0,2) rectangle (axis cs:1,3);
\draw[fill=white, draw=none] (axis cs:1,2) rectangle (axis cs:2,3);
\draw[fill=darkgreen, draw=none] (axis cs:2,2) rectangle (axis cs:3,3);
\draw[fill=darkgreen, draw=none] (axis cs:3,2) rectangle (axis cs:4,3);
\draw[fill=darkgreen, draw=none] (axis cs:4,2) rectangle (axis cs:5,3);
\draw[fill=white, draw=none] (axis cs:5,2) rectangle (axis cs:6,3);
\draw[fill=darkgreen, draw=none] (axis cs:6,2) rectangle (axis cs:7,3);
\draw[fill=darkgreen, draw=none] (axis cs:7,2) rectangle (axis cs:8,3);
\draw[fill=darkgreen, draw=none] (axis cs:8,2) rectangle (axis cs:9,3);
\draw[fill=white, draw=none] (axis cs:9,2) rectangle (axis cs:10,3);
\draw[fill=darkgreen, draw=none] (axis cs:0,3) rectangle (axis cs:1,4);
\draw[fill=darkgreen, draw=none] (axis cs:1,3) rectangle (axis cs:2,4);
\draw[fill=darkgreen, draw=none] (axis cs:2,3) rectangle (axis cs:3,4);
\draw[fill=darkgreen, draw=none] (axis cs:3,3) rectangle (axis cs:4,4);
\draw[fill=darkgreen, draw=none] (axis cs:4,3) rectangle (axis cs:5,4);
\draw[fill=white, draw=none] (axis cs:5,3) rectangle (axis cs:6,4);
\draw[fill=white, draw=none] (axis cs:6,3) rectangle (axis cs:7,4);
\draw[fill=white, draw=none] (axis cs:7,3) rectangle (axis cs:8,4);
\draw[fill=white, draw=none] (axis cs:8,3) rectangle (axis cs:9,4);
\draw[fill=white, draw=none] (axis cs:9,3) rectangle (axis cs:10,4);
\draw[fill=darkgreen, draw=none] (axis cs:0,4) rectangle (axis cs:1,5);
\draw[fill=darkgreen, draw=none] (axis cs:1,4) rectangle (axis cs:2,5);
\draw[fill=darkgreen, draw=none] (axis cs:2,4) rectangle (axis cs:3,5);
\draw[fill=darkgreen, draw=none] (axis cs:3,4) rectangle (axis cs:4,5);
\draw[fill=darkgreen, draw=none] (axis cs:4,4) rectangle (axis cs:5,5);
\draw[fill=darkgreen, draw=none] (axis cs:5,4) rectangle (axis cs:6,5);
\draw[fill=darkgreen, draw=none] (axis cs:6,4) rectangle (axis cs:7,5);
\draw[fill=darkgreen, draw=none] (axis cs:7,4) rectangle (axis cs:8,5);
\draw[fill=darkgreen, draw=none] (axis cs:8,4) rectangle (axis cs:9,5);
\draw[fill=darkgreen, draw=none] (axis cs:9,4) rectangle (axis cs:10,5);
\draw[fill=white, draw=none] (axis cs:0,5) rectangle (axis cs:1,6);
\draw[fill=white, draw=none] (axis cs:1,5) rectangle (axis cs:2,6);
\draw[fill=white, draw=none] (axis cs:2,5) rectangle (axis cs:3,6);
\draw[fill=white, draw=none] (axis cs:3,5) rectangle (axis cs:4,6);
\draw[fill=darkgreen, draw=none] (axis cs:4,5) rectangle (axis cs:5,6);
\draw[fill=darkgreen, draw=none] (axis cs:5,5) rectangle (axis cs:6,6);
\draw[fill=white, draw=none] (axis cs:6,5) rectangle (axis cs:7,6);
\draw[fill=white, draw=none] (axis cs:7,5) rectangle (axis cs:8,6);
\draw[fill=white, draw=none] (axis cs:8,5) rectangle (axis cs:9,6);
\draw[fill=white, draw=none] (axis cs:9,5) rectangle (axis cs:10,6);
\draw[fill=darkgreen, draw=none] (axis cs:0,6) rectangle (axis cs:1,7);
\draw[fill=darkgreen, draw=none] (axis cs:1,6) rectangle (axis cs:2,7);
\draw[fill=darkgreen, draw=none] (axis cs:2,6) rectangle (axis cs:3,7);
\draw[fill=white, draw=none] (axis cs:3,6) rectangle (axis cs:4,7);
\draw[fill=darkgreen, draw=none] (axis cs:4,6) rectangle (axis cs:5,7);
\draw[fill=white, draw=none] (axis cs:5,6) rectangle (axis cs:6,7);
\draw[fill=darkgreen, draw=none] (axis cs:6,6) rectangle (axis cs:7,7);
\draw[fill=white, draw=none] (axis cs:7,6) rectangle (axis cs:8,7);
\draw[fill=white, draw=none] (axis cs:8,6) rectangle (axis cs:9,7);
\draw[fill=white, draw=none] (axis cs:9,6) rectangle (axis cs:10,7);
\draw[fill=darkgreen, draw=none] (axis cs:0,7) rectangle (axis cs:1,8);
\draw[fill=darkgreen, draw=none] (axis cs:1,7) rectangle (axis cs:2,8);
\draw[fill=darkgreen, draw=none] (axis cs:2,7) rectangle (axis cs:3,8);
\draw[fill=white, draw=none] (axis cs:3,7) rectangle (axis cs:4,8);
\draw[fill=darkgreen, draw=none] (axis cs:4,7) rectangle (axis cs:5,8);
\draw[fill=white, draw=none] (axis cs:5,7) rectangle (axis cs:6,8);
\draw[fill=white, draw=none] (axis cs:6,7) rectangle (axis cs:7,8);
\draw[fill=darkgreen, draw=none] (axis cs:7,7) rectangle (axis cs:8,8);
\draw[fill=white, draw=none] (axis cs:8,7) rectangle (axis cs:9,8);
\draw[fill=white, draw=none] (axis cs:9,7) rectangle (axis cs:10,8);
\draw[fill=darkgreen, draw=none] (axis cs:0,8) rectangle (axis cs:1,9);
\draw[fill=white, draw=none] (axis cs:1,8) rectangle (axis cs:2,9);
\draw[fill=darkgreen, draw=none] (axis cs:2,8) rectangle (axis cs:3,9);
\draw[fill=white, draw=none] (axis cs:3,8) rectangle (axis cs:4,9);
\draw[fill=darkgreen, draw=none] (axis cs:4,8) rectangle (axis cs:5,9);
\draw[fill=white, draw=none] (axis cs:5,8) rectangle (axis cs:6,9);
\draw[fill=white, draw=none] (axis cs:6,8) rectangle (axis cs:7,9);
\draw[fill=white, draw=none] (axis cs:7,8) rectangle (axis cs:8,9);
\draw[fill=darkgreen, draw=none] (axis cs:8,8) rectangle (axis cs:9,9);
\draw[fill=white, draw=none] (axis cs:9,8) rectangle (axis cs:10,9);
\draw[fill=white, draw=none] (axis cs:0,9) rectangle (axis cs:1,10);
\draw[fill=white, draw=none] (axis cs:1,9) rectangle (axis cs:2,10);
\draw[fill=white, draw=none] (axis cs:2,9) rectangle (axis cs:3,10);
\draw[fill=white, draw=none] (axis cs:3,9) rectangle (axis cs:4,10);
\draw[fill=darkgreen, draw=none] (axis cs:4,9) rectangle (axis cs:5,10);
\draw[fill=white, draw=none] (axis cs:5,9) rectangle (axis cs:6,10);
\draw[fill=white, draw=none] (axis cs:6,9) rectangle (axis cs:7,10);
\draw[fill=white, draw=none] (axis cs:7,9) rectangle (axis cs:8,10);
\draw[fill=white, draw=none] (axis cs:8,9) rectangle (axis cs:9,10);
\draw[fill=darkgreen, draw=none] (axis cs:9,9) rectangle (axis cs:10,10);
\end{axis}
\end{tikzpicture}%
}
    \end{subfigure}
    \vspace{-8pt}
    \caption{Block structure of $\bK$ when ${\sum}_{i{,}j}\card(\norm{\bK_{ij}}_F){=}50$; Green blocks denote nonzero blocks of $\bK$. From left to right, the subfigures correspond to the results of the proposed algorithm with the weighted-$\ell_1$ norm and cardinality, \citep{lin2013design}, and \citep{lian2018sparsity}.} \label{fig:structure_analysis}
    \vspace*{-0.5\baselineskip} 
\end{figure}
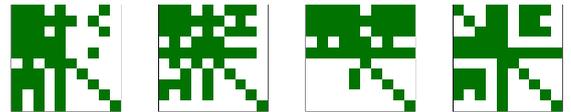


The results in \autoref{fig:performance_analysis}, \autoref{fig:pole_analysis}, and \autoref{fig:structure_analysis} show the \gls{lq} performance, stability, and sparse structure of the network controllers resulting from the proposed method and reference approaches in \citep{lin2013design,lian2018sparsity}.
In all cases, the sparsity-promotion method converges to a sparse controller structure.
The approach in \citep{lin2013design} achieves the best \gls{lq} performance, but can lead to instability due to the agent uncertainty it neglects. 
Both versions of the proposed method, as well as the approach in \citep{lian2018sparsity}, maintain stability under uncertain dynamics and for all vertex combinations of the polytopic uncertainty, with the proposed methods exhibiting superior \gls{lq} performance. 

In practical scenarios such as wire-based communication, enforcing a symmetric controller structure is beneficial for reducing infrastructure requirements.
Moreover, the system matrix $\bA$ in this example has a symmetric network with spatially decaying property, suggesting that a symmetric feedback controller is likely to be advantageous \citep{motee2008optimal}.
As shown in \autoref{fig:structure_analysis}, the proposed method with the cardinality penalty yields a symmetric sparse structure.
These results demonstrate that the proposed methods achieve favorable \gls{lq} performance while preserving stability under system uncertainties using dissipativity, and can offer additional advantages in constructing practical communication architectures.

\begin{table}
\begin{center}
\caption{Computation time when $N=50$}\label{tb:comp_time}
    \vspace{-6pt}
\begin{tabular}{ C{0.05cm} C{1.3cm} C{1.2cm} C{1.9cm} C{2.cm}}
\hline\hline
& \hspace{-10pt}Weighted $\ell_1$ & \hspace{-8pt}Cardinality & \hspace{-8pt}\citep{lin2013design} & \hspace{-8pt}\citep{lian2018sparsity} \\ 
& \hspace{-10pt}8469 s & \hspace{-8pt}17276 s & \hspace{-8pt}4429 s &  \hspace{-8pt} 245669 s \\ \hline\hline
\end{tabular}
\end{center}
\end{table}

The total computation time is also evaluated for a larger network with $N{=}50$ as in \autoref{tb:comp_time}. 
In this case, $\cH_\infty$ constraint parameter $\gamma$ is set to $8$.
Among the stabilizable algorithms, the proposed method with weighted $\ell_1$ norm yields the shortest computation time, as it does not require any iterative procedure to determine a sparse communication topology.



\section{Conclusion}
This paper addressed the problem of dissipativity-based controller synthesis for networks of heterogeneous nonlinear agents, with the goal of identifying an appropriate communication structure. 
Since the original problem is NP hard, a combination of \gls{admm} and \gls{ico} was employed to obtain a solution in a computationally tractable time.
The applicability of the proposed strategy was demonstrated through a numerical example involving a network of both stable and unstable agents exhibiting uncertain dynamical behavior.
The simulation results show that the proposed algorithms can identify network controllers with sparse and practically meaningful communication topologies while achieving moderate \gls{lq} performance and ensuring closed-loop stability under uncertainty. 
The current work focuses on full-state feedback controllers. 
However, the proposed strategy can be extended to dynamic output-feedback settings, which will be addressed in future work.

\bibliography{MyBib}             

\begin{thebibliography}{28}
\providecommand{\natexlab}[1]{#1}
\providecommand{\url}[1]{\texttt{#1}}
\providecommand{\urlprefix}{URL }
\expandafter\ifx\csname urlstyle\endcsname\relax
  \providecommand{\doi}[1]{doi:\discretionary{}{}{}#1}\else
  \providecommand{\doi}{doi:\discretionary{}{}{}\begingroup \urlstyle{rm}\Url}\fi

\bibitem[{Arcak et~al.(2016)Arcak, Meissen, and Packard}]{arcak2016networks}
Arcak, M., Meissen, C., and Packard, A. (2016).
\newblock \emph{Networks of dissipative systems: compositional certification of stability, performance, and safety}.
\newblock Springer.

\bibitem[{Babazadeh and Nobakhti(2016)}]{babazadeh2016sparsity}
Babazadeh, M. and Nobakhti, A. (2016).
\newblock Sparsity promotion in state feedback controller design.
\newblock \emph{{IEEE} Tr. Aut. Ctrl.}, 62(8), 4066--4072.

\bibitem[{Boyd et~al.(2011)Boyd, Parikh, Chu, Peleato, Eckstein et~al.}]{boyd2011distributed}
Boyd, S., Parikh, N., Chu, E., Peleato, B., Eckstein, J., et~al. (2011).
\newblock Distributed optimization and statistical learning via the alternating direction method of multipliers.
\newblock \emph{F. T. Mach.}, 3(1), 1--122.

\bibitem[{Bridgeman and Forbes(2016)}]{bridgeman2016conic}
Bridgeman, L.J. and Forbes, J.R. (2016).
\newblock Conic bounds for systems subject to delays.
\newblock \emph{{IEEE} Tr. Aut. Ctrl.}, 62(4), 2006--2013.

\bibitem[{Fardad et~al.(2011)Fardad, Lin, and Jovanovi{\'c}}]{fardad2011sparsity}
Fardad, M., Lin, F., and Jovanovi{\'c}, M.R. (2011).
\newblock Sparsity-promoting optimal control for a class of distributed systems.
\newblock In \emph{Proc. Amer. Ctrl. Conf.}, 2050--2055. IEEE.

\bibitem[{Gupta(1996)}]{gupta1996robust}
Gupta, S. (1996).
\newblock Robust stabilization of uncertain systems based on energy dissipation concepts.
\newblock Technical report, NASA.

\bibitem[{Hill and Moylan(2003)}]{hill2003stability}
Hill, D. and Moylan, P. (2003).
\newblock The stability of nonlinear dissipative systems.
\newblock \emph{{IEEE} Tr. Aut. Ctrl.}, 21(5), 708--711.

\bibitem[{Jovanovi{\'c} and Dhingra(2016)}]{jovanovic2016controller}
Jovanovi{\'c}, M.R. and Dhingra, N.K. (2016).
\newblock Controller architectures: Tradeoffs between performance and structure.
\newblock \emph{Eu. J. Ctrl.}, 30, 76--91.

\bibitem[{Khalil et~al.(1996)Khalil, Doyle, and Glover}]{khalil1996robust}
Khalil, I., Doyle, J., and Glover, K. (1996).
\newblock \emph{Robust and optimal control}, volume~2.
\newblock Prentice hall New York.

\bibitem[{Lian et~al.(2017)Lian, Chakrabortty, and Duel-Hallen}]{lian2017game}
Lian, F., Chakrabortty, A., and Duel-Hallen, A. (2017).
\newblock Game-theoretic multi-agent control and network cost allocation under communication constraints.
\newblock \emph{IEEE journal on selected areas in communications}, 35(2), 330--340.

\bibitem[{Lian et~al.(2018)Lian, Chakrabortty, Wu, and Duel-Hallen}]{lian2018sparsity}
Lian, F., Chakrabortty, A., Wu, F., and Duel-Hallen, A. (2018).
\newblock Sparsity-constrained mixed {$H_2/H_\infty$} control.
\newblock In \emph{Proc. Amer. Ctrl. Conf.}, 6253--6258. IEEE.

\bibitem[{Lin et~al.(2013)Lin, Fardad, and Jovanovi{\'c}}]{lin2013design}
Lin, F., Fardad, M., and Jovanovi{\'c}, M.R. (2013).
\newblock Design of optimal sparse feedback gains via the alternating direction method of multipliers.
\newblock \emph{{IEEE} Tr. Aut. Ctrl.}, 58(9), 2426--2431.

\bibitem[{LoCicero and Bridgeman(2022)}]{locicero2022sparsity}
LoCicero, E.J. and Bridgeman, L. (2022).
\newblock Sparsity promoting fixed-order {$H_2$}-conic control.
\newblock In \emph{Proc. Amer. Ctrl. Conf.}, 4862--4867. IEEE.

\bibitem[{LoCicero and Bridgeman(2025)}]{locicero2025dissipativity}
LoCicero, E.J. and Bridgeman, L. (2025).
\newblock Dissipativity-augmented multiobjective control of networks.
\newblock \emph{International Journal of Control}, 1--16.

\bibitem[{Lozano et~al.(2013)Lozano, Brogliato, Egeland, and Maschke}]{lozano2013dissipative}
Lozano, R., Brogliato, B., Egeland, O., and Maschke, B. (2013).
\newblock \emph{Dissipative systems analysis and control: theory and applications}.
\newblock Springer Science \& Business Media.

\bibitem[{Motee and Jadbabaie(2008)}]{motee2008optimal}
Motee, N. and Jadbabaie, A. (2008).
\newblock Optimal control of spatially distributed systems.
\newblock \emph{{IEEE} Tr. Aut. Ctrl.}, 53(7), 1616--1629.

\bibitem[{Moylan and Hill(2003)}]{moylan2003stability}
Moylan, P. and Hill, D. (2003).
\newblock Stability criteria for large-scale systems.
\newblock \emph{{IEEE} Tr. Aut. Ctrl.}, 23(2), 143--149.

\bibitem[{Natarajan(1995)}]{natarajan1995sparse}
Natarajan, B.K. (1995).
\newblock Sparse approximate solutions to linear systems.
\newblock \emph{SIAM journal on computing}, 24(2), 227--234.

\bibitem[{Negi and Chakrabortty(2020)}]{negi2020sparsity}
Negi, N. and Chakrabortty, A. (2020).
\newblock Sparsity-promoting optimal control of cyber--physical systems over shared communication networks.
\newblock \emph{Automatica}, 122, 109217.

\bibitem[{Ren et~al.(2021)Ren, Li, Liu, and Ding}]{ren2021successive}
Ren, Y., Li, Q., Liu, K.Z., and Ding, D.W. (2021).
\newblock A successive convex optimization method for bilinear matrix inequality problems and its application to static output-feedback control.
\newblock \emph{I. J. Robu. Nonl. Ctrl.}, 31(18), 9709--9730.

\bibitem[{Romer et~al.(2017)Romer, Montenbruck, and Allg{\"o}wer}]{romer2017determining}
Romer, A., Montenbruck, J.M., and Allg{\"o}wer, F. (2017).
\newblock Determining dissipation inequalities from input-output samples.
\newblock \emph{IFAC-PapersOnLine}, 50(1), 7789--7794.

\bibitem[{Sebe(2018)}]{sebe2018sequential}
Sebe, N. (2018).
\newblock Sequential convex overbounding approximation method for bilinear matrix inequality problems.
\newblock \emph{IFAC-PapersOnLine}, 51(25), 102--109.

\bibitem[{Strong et~al.(2024)Strong, Lavaei, and Bridgeman}]{Strong2024IterativeGain}
Strong, A.K., Lavaei, R., and Bridgeman, L.J. (2024).
\newblock Improved small-signal $\mathcal{L}_{2}$-gain analysis for nonlinear systems.
\newblock In \emph{2024 American Control Conference (ACC)}, 3377--3382.
\newblock \doi{10.23919/ACC60939.2024.10644731}.

\bibitem[{Vidyasagar(1981)}]{vidyasagar1981input}
Vidyasagar, M. (1981).
\newblock \emph{Input-output analysis of large-scale interconnected systems: decomposition, well-posedness and stability}.
\newblock Springer.

\bibitem[{Walsh and Forbes(2019)}]{walsh2019interior}
Walsh, A. and Forbes, J.R. (2019).
\newblock Interior-conic polytopic systems analysis and control.
\newblock \emph{ArXiv}.

\bibitem[{Warner and Scruggs(2017)}]{warner2017iterative}
Warner, E. and Scruggs, J. (2017).
\newblock Iterative convex overbounding algorithms for bmi optimization problems.
\newblock \emph{IFAC-PapersOnLine}, 50(1), 10449--10455.

\bibitem[{Willems(1972)}]{willems1972dissipative}
Willems, J.C. (1972).
\newblock Dissipative dynamical systems part i: General theory.
\newblock \emph{Ar. rat. mech. analy.}, 45(5), 321--351.

\bibitem[{Zakeri and Antsaklis(2022)}]{zakeri2022passivity}
Zakeri, H. and Antsaklis, P.J. (2022).
\newblock Passivity measures in cyberphysical systems design: An overview of recent results and applications.
\newblock \emph{Ctrl. Sys. M.}, 42(2), 118--130.

\end{thebibliography}
                                                   







\appendix
\end{document}